\documentclass[10pt]{article}
\def\bdk{\begin{description} \itemsep=-\parsep \itemindent=-0.9 cm}
\def\edk{\end{description}}
\def \leaderfill{\leaders\hbox to 0.35em{\hss$\cdot$\hss}\hfill}

\usepackage{mathrsfs}              
\usepackage{amsmath}               
\usepackage{bm}                    
\usepackage{amssymb,color}         
\usepackage{graphicx}              
\usepackage[figuresright]{rotating}
\usepackage{amsthm}                
\usepackage{cases}                 
\usepackage{multicol}              
\usepackage{enumerate}             
\usepackage{extarrows}             
\usepackage{graphicx}              
\usepackage{longtable}             
\usepackage{supertabular}          
\usepackage{color}                
\usepackage{float}
\usepackage{changepage}
\usepackage{lscape}
\usepackage{multirow}
\usepackage[comma]{natbib}
\usepackage[colorlinks,citecolor=blue]{hyperref}
\usepackage[justification=centering]{caption}
\makeatletter
 \renewcommand\@biblabel[1]{}
\renewenvironment{thebibliography}[1]
      {\section*{\refname}%
       \@mkboth{\MakeUppercase\refname}{\MakeUppercase\refname}%
       \list{\@biblabel{\@arabic\c@enumiv}}%
            {\settowidth\labelwidth{\@biblabel{#1}}%
             \leftmargin\labelwidth
             \advance\leftmargin\labelsep
             \advance\leftmargin by 2em%
             \itemindent -2em
             \@openbib@code
             \usecounter{enumiv}%
             \let\p@enumiv\@empty
             \renewcommand\theenumiv{\@arabic\c@enumiv}}%
       \sloppy
       \clubpenalty4000
       \@clubpenalty \clubpenalty
       \widowpenalty4000%
       \sfcode`\.\@m}
      {\def\@noitemerr
        {\@latex@warning{Empty `thebibliography' environment}}%
       \endlist}
 \makeatother

\makeatletter
\newcommand{\rmnum}[1]{\romannumeral #1}                           
\newcommand{\Rmnum}[1]{\expandafter\@slowromancap\romannumeral #1@}
\makeatother

\renewcommand{\theequation}{\arabic{section}.\arabic{equation}}  

\newtheorem{satz1}{Satz1}[section]                       
\newtheorem{satz4}{Satz4}[section]
\newtheorem{satz5}{Satz5}[section]
\newtheorem{satz6}{Satz6}[section]

\newtheorem{thm}[satz1]{ Theorem}  
\newtheorem{proposition}[satz4]{Proposition}
\newtheorem{remark}[satz5]{Remark}
\newtheorem{defn}[satz6]{Definition}

\catcode`@=11 \@addtoreset{equation}{section} \catcode`@=12

\begin{document}
\title{A new First-Order mixture integer-valued threshold autoregressive process based on binomial thinning and negative binomial thinning}
\author{\small Danshu Sheng$^1$,~Dehui Wang$^{1,}$\footnotemark[1]~,~Liuquan Sun$^2$~\\[0.0cm]
{\small\it 1. School of Mathematics and Statistics, Liaoning University, Shenyang, China.}\\
{\small\it 2. Institute of Applied Mathematics, Academy of Mathematics and Systems Science,}\\
{\small\it  Chinese Academy of Sciences, Beijing 100190, China}\\
}
\date{}
\footnotetext[1] {Corresponding author: wangdehui@lnu.edu.cn}

\maketitle
\begin{center}
\begin{minipage}{13.5truecm}
{\bf Abstract}
In this paper, we introduce a new first-order mixture integer-valued threshold autoregressive process, based on the binomial and negative binomial thinning operators. Basic probabilistic and statistical properties of this model are discussed. Conditional least squares (CLS) and conditional maximum likelihood (CML) estimators are derived and the asymptotic properties of the estimators are established. The inference for the threshold parameter is obtained based on the CLS and CML score functions.
Moreover, the Wald test is applied to detect the existence of the piecewise structure. Simulation studies are considered, along with an application: the number of criminal mischief incidents in the Pittsburgh dataset.
\end{minipage}
\end{center}

\section{Introduction}
Threshold time series model and its applications have had a very large influence on various fields of research since the groundbreaking works of \cite{Tong1978,Tong1983}. 
For instance, \cite{Chen2011} provide an overview of the development of threshold autoregression (TAR) models and their use in finance. The use of TAR in economics is mentioned in \cite{Hansen2011}.
\cite{Chan2004} introduced the nonlinear TAR model and applied it to actuarial science for insurance product pricing.
TAR models often involve piecewise linearization by dividing a complicated system into regimes according to some threshold, offering a somewhat simple method of fitting a complex system. In recent years, to capture the piecewise phenomenon of discrete-valued time series, \cite{Wang2014} introduced a self-excited threshold Poisson autoregressive (SETPAR) model and applied it to global major earthquake data. \cite{Moller2016} proposed a basic self-exciting threshold binomial AR(1) model (SETBAR(1)) with values across a finite range of counts.
\cite{MW2015} presented a brief survey of threshold models for integer-valued time series with an infinite range and introduced two new models for the case of a finite range.
Additionally, some academics have studied the following self-excited threshold integer-valued autoregressive (SETINAR) types models based on different thinning operators,
\[X_t = \left\{ \begin{array}{ll}
\alpha_1\circ X_{t-1}+Z_t,&X_{t-1}\leq r\\
\alpha_2\circ X_{t-1}+Z_t,& X_{t-1}> r;
\end{array}\right.\]
where $\{Z_t\}$ is a sequence of i.i.d. random variables. For example, based on the binomial thinning operator (``$\circ$"), \cite{Monteiro2012} presented an integer-valued self-exciting threshold autoregressive (SETINAR(2,1)) process, and  based on the negative binomial thinning operator (``$\ast$"), \cite{Yang2018} studied an integer-valued TAR procedure (NBTINAR(1)). The definitions of the binomial thinning operator ``$\circ$" and the negative binomial thinning operator ``$\ast$" are defined in Definition \ref{BiNB}-(ii) and (iii).

It is clear that when the integer-valued time series model is defined using thinning operators,
the operator characteristics are important because they affect the statistical properties of the model.
As introduced by \cite{Na2017}, the negative binomial thinning operators, might be used to describe elements or random events that, by self-replication, influence other elements, cause other random events, or otherwise contribute to the overall thinning total by more than 1. Thus, the random number produced by $\alpha\ast X_{t-1}$ might theoretically be larger than $X_{t-1}$, and the ``$\ast$" operator is better suited for fitting time series data with significant volatility. Furthermore, in the case of the binomial thinning operator, since it is the sum of $X_{t-1}$ Bernoulli random variables, it cannot generate a random number larger than $X_{t-1}$, which means that the binomial thinning operator's contribution to $X_t$ is less than $X_{t-1}$.

Note that the SETINAR models described above choose the same operator for both segments, but in practice, the operator may vary depending on the value of $X_{t-1}$.
That is, when $X_{t-1}$ is in a range, it contributes to $X_t$ in the form of a binomial thinning operator.
However, when it is not in this range, it may contribute to $X_t$ as a negative binomial thinning operator.
Precisely for this reason, herein, we define a new thinning operator as a mixture of the binomial and the negative binomial operator.
Moreover, the mixing is self-excited, so it is very capable of handling the requirements of certain counting data. Likewise, there are a variety of real-life examples where these types of mixed thinning integer-valued threshold autoregressive (TINAR) models are more likely to be used than the TINAR models based only on one of these two thinning operators.

Consider the number of infectious illness cases that are confirmed each day.
The following will happen if the infectious disease being studied has obvious disease characteristics, has a significant impact on people, and has an incubation period that is shorter than the counting interval, where the incubation period is the period of time between exposure to pathogenic microorganisms and the onset of obvious symptoms.
The patient will be kept completely alone in the hospital, only being able to communicate with the attending physician.
In this situation, it makes sense to explain the contribution of the number of confirmed cases at time $t-1$ ($X_{t-1}$) to the number of confirmed cases at time $t$ ($X_t$) using the INAR(1) model based on the binomial thinning operator (Bi-INAR(1)). However, if there are more than a certain number of confirmed cases, the hospital will be unable to treat everyone, forcing the patients to self-isolate at home.
Patients will likely interact with many individuals as a result of this loose isolation, which will increase infections and the transmission rate of the infectious disease. An INAR(1) model based on the negative binomial thinning operator (NB-INAR(1)) can be used to characterize the number of confirmed cases at time $t$.

Of course, there are also cases where the characteristics of the infectious diseases studied are not obvious, and when the number of patients is small, it may often go unnoticed. At this time, the infection tends to display a cross-infection pattern, and the number of infectious diseases can be described by the NB-INAR(1) model. However, with an increase in the number of confirmed cases, national or regional governments will pay greater attention, and they will develop a series of plans to reduce this cross-infection; thus, it is more reasonable to use the Bi-INAR(1) model to describe the number of patients.

Motivated by the aforementioned examples, we propose a new first-order mixture thinning (binomial thinning and negative binomial thinning) integer-valued threshold autoregressive (BiNB-MTTINAR(1)) process. For this, we initially give the definition of the BiNB-MTTINAR(1) model and study the statistical inference for the proposed model. Furthermore, considering that the model will appear different from the general SETINAR model when $\alpha_1=\alpha_2$, we propose a new method to estimate the threshold parameter $r$, and present a new Wald test for conditional variance to detect the existence of the piecewise structure. Finally, from the application, we can also see that our proposed model is very competitive.

The paper is organized as follows:
In Section 2, we introduce the BiNB-MTINAR(1) process and discuss its basic probabilistic and statistical properties.
In Section 3, we propose two estimation methods for estimating the model parameters and threshold value.
We construct two Wald test statistics for conditional expectation and conditional variance parameters, respectively, to test the existence of the piecewise structure.
In Section 4, some simulation results for the estimation methods, the size and power of Wald tests are presented.
Real data example is given in Section 5.
Some concluding remarks are given in Section 6.
All proofs are postponed to the Appendix.

\section{The BiNB-MTTINAR(1) model}
We first introduce the definition of the BiNB-MTTINAR(1) process.
\begin{defn}\label{BiNB}
The process $\{X_t\}$ is called BiNB-MTTINAR(1) process if $X_t$ follows the recursion
\begin{align}\label{R0}
X_t = \left\{ \begin{array}{ll}
\phi_1\circ X_{t-1}+Z_{1,t},&X_{t-1}\leq r\\
\phi_2\ast X_{t-1}+Z_{2,t},& X_{t-1}> r,
\end{array}\right.
\end{align}
or
\begin{align}\label{R1}
X_t =\left\{ \begin{array}{ll}
\phi_2\ast X_{t-1}+Z_{2,t},&X_{t-1}\leq r\\
\phi_1\circ X_{t-1}+Z_{1,t},& X_{t-1}> r.
\end{array} \right.
\end{align}
For convenience, we write the above two models by the symbol $R$ as follows
\begin{equation}\label{Mymodel}
  X_t=(\phi_{1}\circ X_{t-1}+Z_{1,t})I_{1,t}^R+(\phi_{2}\ast X_{t-1}+Z_{2,t})I_{2,t}^R,~~t \in \mathbb{Z},
\end{equation}
where
\begin{enumerate}[(i)]
 \setlength{\itemsep}{0pt}
 \setlength{\parskip}{0pt}
 \setlength{\parsep}{0pt}
\item
$
I_{1,t}^R=\left\{\begin{array}{l}
I\{{X_{t-1} \leq r}\},R=0,\\
I\{{X_{t-1} > r}\},R=1,
\end{array} \right.
$
and
$I_{2,t}^R=1-I_{1,t}^R=\left\{\begin{array}{l}
I\{{X_{t-1} > r}\},R=0,\\
I\{{X_{t-1} \leq r}\},R=1;
\end{array} \right.$
That is, $R=0$ indicates that BiNB-MTTINAR(1) represents the process (\ref{R0}).
  \item the binomial thinning operator ``$\phi_{1}\circ$", proposed by \cite{Steutel1979}, is defined as $\phi_1\circ X=\sum_{i=1}^X B_i$, where $\phi_1 \in (0,1)$, $\{B_i\}$ is a sequence of i.i.d. Bernoulli random variables satisfying $P(B_i=1)=1-P(B_i=0)=\phi_1$, $B_i$ is independent of $X$;

  \item the negative binomial thinning operator ``$\phi_{2}\ast$", proposed by \cite{Ristic2009}, is defined as $\phi_2\ast X=\sum_{i=1}^X W_i$, where $\phi_2 \in (0,1)$, $\{W_i\}$ is a sequence of i.i.d. Geometric random variables with parameter $\frac{\phi_2}{1+\phi_2}$, $W_i$ is independent of $X$;
  \item $\{Z_{1,t}\}$ is a sequence of i.i.d. Poisson distributed random variables with mean $\lambda$, $\{Z_{2,t}\}$ is a sequence of i.i.d. geometric distributed random variables with parameter $\frac{\lambda}{1+\lambda}$, $\lambda\in(0,\infty)$ ;
  \item For fixed $t$, $Z_{1,t}$ is assumed to be independent of $\phi_{1}\circ X_{t-1}$ and $X_{t-l}$ for all $l{\geq}1$, $Z_{2,t}$ is assumed to be independent of $\phi_{2}\ast X_{t-1}$ and $X_{t-l}$ for all $l{\geq}1$.
\end{enumerate}
\end{defn}
Note that compared with the traditional integer-valued threshold models, the condition $\phi_1\neq\phi_2$ is not necessary, which can reflect a scenario where the variance of the two regimes is different but the mean is the same. We can give an example to illustrate this situation.
For instance, viral data show that for some viruses, their own activity remains unaltered, which means that their probability of transmission is unaltered. This may also be regarded as their conditional expectation of transmission being unchanged.  However, due to the increase in the number of patients, resulting in a series of external influences such as cross-infection and increased virus density in the air, the virus spread faster, that is, the volatility (conditional variance) increased. The BiNB-MTTINAR(1) model is more suited for simulation at this time.
In the following Remark, how to select $R$ in practice is discussed.
\begin{remark}
In practice, we are more likely to select the BiNB-MTTINAR(1) model with $R=0$ if the variance difference between the two regimes is extremely large. This is due to the fact that the variance difference between the two regimes at $R=0$ is greater than it is at $R=1$ based on the definitions of thinning operators and  Poisson, Geometric random variables.
\end{remark}
Obviously, the BiNB-MTTINAR(1) model is Markovian, transition probabilities are very important when we consider the conditional maximum likelihood estimate for the proposed model. So we first give the transition probabilities of the BiNB-MTTINAR(1) process as follows,
\begin{equation}\label{transprob}
\begin{split}
P(i,j)&=P(X_t=j|X_{t-1}=i)\\
&=P\left((\phi_{1}\circ X_{t-1}+Z_{1,t})I_{1,t}^R+(\phi_{2}\ast X_{t-1}+Z_{2,t})I_{2,t}^R=j|X_{t-1}=i\right)\\
&=p_1(i,j,\phi_{1},\lambda)I_{1,t}^R+p_2(i,j,\phi_{2},\lambda)I_{2,t}^R,
\end{split}
\end{equation}
\vspace{1cm}
where
\begin{equation}\label{zygl}
\begin{split}
p_1(i,j,\phi_{1},\lambda)&=\sum_{m=0}^{\min(i,j)}
\binom{i}{m}
e^{-\lambda}\frac{\lambda^{j-m}}{(j-m)!}
\phi_1^m(1-\phi_1)^{i-m},\\
p_2(i,j,\phi_{2},\lambda)&=\sum_{m=0}^{j}
\frac{\Gamma(i+m)}{\Gamma(i)\Gamma(m+1)}\frac{\phi_2^m}{(1+\phi_2)^{i+m}}\frac{\lambda^{j-m}}{(1+\lambda)^{j-m+1}}.
\end{split}
\end{equation}

Next, we are ready to state that there exists a strictly stationary and ergodicity of {\color{red}the} BiNB-MTTINAR(1) process satisfying Definition \ref{BiNB}.
Proofs of the following propositions are in the Appendix.
\begin{proposition}\label{stationarity}
Let $\{X_t\}_{t \in \mathbb{Z}}$ be the process defined in Definition \ref{BiNB}.
Then $\{X_t\}_{t \in \mathbb{Z}}$ is an irreducible, aperiodic, and positive recurrent (and hence ergodic) Markov chain. Thus, there exists a strictly stationary process satisfying (\ref{Mymodel}).
\end{proposition}
Since the existence of the first three moments is a necessary condition for deriving the asymptotic properties of the parameter estimation in Sect. 3, and the moments and conditional moments are useful in obtaining the appropriate estimating equations for parameter estimation, we then give the following propositions. For simplicity in notation, we denote ${\rm E}(I_{1,t}^R)=p_1=q$, ${\rm E}(I_{2,t}^R)=p_2=1-q$, $\mu_{1}:={\rm E}(X_{t}|X_{t}\leq r)$, $\mu_{2}:={\rm E}(X_{t}|X_{t}>r)$,
$\sigma_{1}^{2}:={\rm Var}(X_{t}|X_{t}\leq r)$, $\sigma_{2}^{2}:={\rm Var}(X_{t}|X_{t}>r)$,
$\gamma_h^{(1)}:={\rm Cov}(X_{t},X_{t+h}|X_{t+h}\leq r)$, $\gamma_h^{(2)}:={\rm Cov}(X_{t},X_{t+h}|X_{t+h}>r),$
where $\gamma_{0}^{(s)}=[(\sigma_s^2+\mu_s^2)-\mu_s{\rm E}(X_t)],s=1,2.$
\begin{proposition}\label{momentexist}
Let $\{X_t\}$ be the process defined by Definition \ref{BiNB}. Then ${\rm E}(X_t^k) < \infty$ for $k = 1,2,3.$
\end{proposition}

\begin{proposition}\label{momentform}
Let $\{X_t\}$ be the process defined by Definition \ref{BiNB}. Then
\begin{enumerate}[$($i$)$]
 \setlength{\itemsep}{0pt}
 \setlength{\parskip}{0pt}
 \setlength{\parsep}{0pt}
\item ${\rm E}(X_t|X_{t-1})=\phi_{1}X_{t-1}I_{1,t}^R+\phi_{2}X_{t-1}I_{2,t}^R+\lambda$;
\item ${\rm Var}(X_t|X_{t-1})=(\phi_{1}(1-\phi_{1})X_{t-1}+\lambda)I_{1,t}^R+(\phi_{2}(1+\phi_2)X_{t-1}+\lambda(1+\lambda))I_{2,t}^R$;
\item ${\rm E}(X_t)=q\phi_{1}\mu_{1}+(1-q)\phi_{2}\mu_{2}+\lambda$;
\item ${\rm Var}(X_t)= q\left[\phi_1^2\sigma_1^2+\phi_1(1-\phi_1)\mu_1\right]+q(1-q)\phi_1^2\mu_1^2+q\lambda
    +(1-q)\left[\phi_2^2\sigma_2^2+\phi_2(1+\phi_2)\mu_2\right]\protect\\
    ~~~~~~~~~~~+q(1-q)\phi_2^2\mu_2^2+(1-q)\lambda(1+\lambda)-2q(1-q)(\phi_1\mu_1+\lambda)(\phi_2\mu_2+\lambda)$;
\item ${\rm Cov}(X_{t},X_{t+h})=\sum_{s=1}^2\phi_sp_s\gamma_{h-1}^{(s)}$;
\item $\rho(h):={\rm Corr}(X_{t},X_{t+h})=[\sum_{s=1}^2\phi_sp_s\gamma_{h-1}^{(s)}]\setminus{\rm Var}(X_t) $.
\end{enumerate}
\end{proposition}
\section{Parameters Estimation for the BiNB-MTTINAR(1) model}
Suppose we have a series of observations $\{X_t\}_{t=1}^n$ generated from the BiNB-MTTINAR(1) process. We first estimate the parameter $\bm{\theta}=(\phi_{1},\phi_{2},\lambda)^\textsf{T}$ using conditional least squares (CLS) and conditional maximum likelihood (CML) methods, the threshold parameter $r$ is assumed to be known until Sect. 3.3 where we will propose two methods to handle the unknown $r$ case. Later in Sect. 3.4,
we will provide two Wald test statistics for conditional expectation and conditional variance parameters to check the nonlinearity of the data. All the proofs are presented in the Appendix.
\subsection{CLS Estimation}
Let $g(\bm{\theta},X_{t-1})={\rm E}(X_t|X_{t-1})=\phi_1 X_{t-1}I_{1,t}^R+\phi_2 X_{t-1}I_{2,t}^R+\lambda$,
then the CLS estimator $\hat{\bm{\theta}}_{CLS}:=(\hat{\phi}_{1,CLS},\hat{\phi}_{2,CLS},\hat{\lambda}_{CLS})^\textsf{T}$
of $\bm{\theta}$ is obtained by minimizing the sum of the squared deviations
\begin{align}
Q(\bm{\theta})&:=\sum_{t=1}^n(X_t-g(\bm{\theta},X_{t-1}))^2=\sum_{t=1}^n U_t^2(\bm{\theta}),\nonumber
\end{align}
where
$$
U_t(\bm{\theta})=X_t-\phi_1 X_{t-1}I_{1,t}^R-\phi_2 X_{t-1}I_{2,t}^R-\lambda.
$$
From he first partial derivative equal to $0$, after some calculations, we obtain the CLS estimators  closed-form expressions as follows
\begin{equation}\label{CLS_eq}
\begin{split}
\hat{\phi}_{1,CLS}&=\frac{\left(n\sum\limits_{t=1}^{n}I_{2,t}^RX_{t-1}^2-(\sum\limits_{t=1}^{n}I_{2,t}^RX_{t-1})^2\right)M_1+\sum\limits_{t=1}^{n}I_{1,t}^RX_{t-1}\left(\sum\limits_{t=1}^{n}I_{2,t}^RX_{t-1}M_2-\sum\limits_{t=1}^{n}I_{2,t}^RX_{t-1}^2M_3\right)}
{n\sum\limits_{k=1}^{2}\left(\sum\limits_{t=1}^{n}I_{k,t}^RX_{t-1}^2\right)-(\sum\limits_{t=1}^{n}I_{1,t}^RX_{t-1})^2\sum\limits_{t=1}^{n}I_{2,t}^RX_{t-1}^2-(\sum\limits_{t=1}^{n}I_{2,t}^RX_{t-1})^2\sum\limits_{t=1}^{n}I_{1,t}^RX_{t-1}^2},\\
\hat{\phi}_{2,CLS}&=\frac{(n\sum\limits_{t=1}^{n}I_{1,t}^RX_{t-1}^2-(\sum\limits_{t=1}^{n}I_{1,t}^RX_{t-1})^2)M_2+\sum\limits_{t=1}^{n}I_{2,t}^RX_{t-1}\left(\sum\limits_{t=1}^{n}I_{1,t}^RX_{t-1}M_1-\sum\limits_{t=1}^{n}I_{1,t}^RX_{t-1}^2M_3\right)}
{n\sum\limits_{k=1}^{2}\left(\sum\limits_{t=1}^{n}I_{k,t}^RX_{t-1}^2\right)-(\sum\limits_{t=1}^{n}I_{1,t}^RX_{t-1})^2\sum\limits_{t=1}^{n}I_{2,t}^RX_{t-1}^2-(\sum\limits_{t=1}^{n}I_{2,t}^RX_{t-1})^2\sum\limits_{t=1}^{n}I_{1,t}^RX_{t-1}^2},\\
\hat{\lambda}_{CLS}&=\frac{\sum\limits_{t=1}^{n}I_{1,t}^RX_{t-1}^2\left(\sum\limits_{t=1}^{n}I_{2,t}^RX_{t-1}^2M_3-\sum\limits_{t=1}^{n}I_{2,t}^RX_{t-1}M_2\right)-\sum\limits_{t=1}^{n}I_{1,t}^RX_{t-1}\sum\limits_{t=1}^{n}I_{2,t}^RX_{t-1}^2M_1}
{n\sum\limits_{k=1}^{2}\left(\sum\limits_{t=1}^{n}I_{k,t}^RX_{t-1}^2\right)-(\sum\limits_{t=1}^{n}I_{1,t}^RX_{t-1})^2\sum\limits_{t=1}^{n}I_{2,t}^RX_{t-1}^2-(\sum\limits_{t=1}^{n}I_{2,t}^RX_{t-1})^2\sum\limits_{t=1}^{n}I_{1,t}^RX_{t-1}^2},
\end{split}
\end{equation}
where
\begin{align*}
M_1=\sum_{t=1}^n X_{t}X_{t-1}I_{1,t}^R,~~~M_2=\sum_{t=1}^{n}X_{t}X_{t-1}I_{2,t}^R,~~~M_3=\sum_{t=1}^{n}X_{t}.
\end{align*}

From Propositions \ref{stationarity} and \ref{momentexist}, the BiNB-MTTINAR(1) model is stationary, ergodic, and the first three moments are bounded, it follows from Theorems 3.1 and 3.2 in \cite{Klimko1978} that $\hat{\bm{\theta}}_{CLS}$ is a consistent and asymptotically normal estimator of $\bm{\theta}$.
That is, the following theorem about the consistency and asymptotic properties of the CLS estimator is valid.
\begin{thm}\label{clsnormal}
Let $\{X_t\}$ be a BiNB-MTTINAR(1) process. Then the CLS estimator $\hat{\bm{\theta}}_{CLS}$ is consistent and has the asymptotically distribution,
\begin{align}\label{clsnormal_eq}
\sqrt{n}(\hat{\bm{\theta}}_{CLS}-\bm{\theta_0}) \overset{d}{\longrightarrow}N(\bm{0},\bm{V}^{-1}\bm{W}\bm{V}^{-1}),
\end{align}
where $\bm{V}$ and $\bm{W}$ are square matrices of order 3 with the $ij$th element given by
$$V_{ij}:={\rm E}\left(\frac{\partial}{\partial{\bm{\theta}}_{i}}g(\bm{\theta},X_{t-1})\frac{\partial}{\partial\bm{\theta}_{j}}g(\bm{\theta},X_{t-1})\right)_{\bm{\theta_0}},$$
$$W_{ij}:={\rm E}\left(U_{t}^2(\bm{\theta})\frac{\partial}{\partial\bm{\theta}_i}g(\bm{\theta},X_{t-1})\frac{\partial}{\partial\bm{\theta}_j}g(\bm{\theta},X_{t-1})\right)_{\bm{\theta_0}}.$$
\end{thm}
\subsection{CML Estimation}
For a fixed value of $x_0$, the conditional likelihood function for the BiNB-MTTINAR(1) model can be written as
\begin{align*}
L(\bm{\theta}):=P(X_1=x_1,\cdots,X_n=x_n|x_0)=\prod^n_{t=1}P(x_{t-1},x_t),
\end{align*}
where $P(x_{t-1},x_t)$ is the transition probabilities defined in (\ref{transprob}). The CML estimator $\hat{\bm{\theta}}_{CML}:=(\hat{\phi}_{1,CML},\hat{\phi}_{2,CML},\hat{\lambda}_{CML})^\textsf{T}$
of $\bm{\theta}$ is obtained by maximizing the conditional log-likelihood function
\begin{align*}
\ell(\bm{\theta})&=\log L(\bm{\theta})=\sum_{t=1}^n\ell_t(\bm{\theta}),
\end{align*}
where $\ell_t(\bm{\theta})=\log P(x_{t-1},x_t).$

Although  no closed-form expressions for the estimators can be found, the numerical solutions can be solved by the MATLAB function \verb"fmincon" or the R Function \verb"optim" and the initial estimators required are obtained by the CLS estimators (\ref{CLS_eq}). The details of the numerical process are discussed later in Sect. 4. Next, we give the following theorem to state the consistency and asymptotic properties of the CML estimator.
\begin{thm}\label{cmlzt}
Let $\{X_t\}$ be a BiNB-MTTINAR(1) process, then the CML estimator $\hat{\bm{\theta}}_{CML}$ is consistent and has the following asymptotically distribution,
\begin{align*}
\sqrt{n}(\hat{\bm{\theta}}_{CML}-\bm{\theta_0})\overset{d}{\longrightarrow}N(\bm{0},\bm{J}^{-1}(\bm{\theta_0})\bm{I}(\bm{\theta_0})\bm{J}^{-1}(\bm{\theta_0})),
\end{align*}
where $\bm{I}(\bm{\theta_0})=\mathrm{E}\left[\frac{\partial l_t(\bm{\theta})}{\partial \bm{\theta}}\frac{\partial l_t(\bm{\theta})}{\partial \bm{\theta}^{\mathrm{T}}}\right]_{\bm{\theta_0}}$, $\bm{J}(\bm{\theta_0})=\mathrm{E}\left[\frac{\partial^2 l_t(\bm{\theta})}{\partial \bm{\theta}{\partial \bm{\theta}^{\mathrm{T}}}}\right]_{\bm{\theta_0}}$.
\end{thm}
\subsection{Inference Method for Threshold $r$}
In this section, we turn to the estimation of the threshold parameter $r$. Since $r$ is an integer, the CLS and CML score functions are not differentiable with respect  to $r$.
As a result, the methods for selecting the threshold $r$ for integer-valued threshold models are primarily to search for the optimal value within a fixed interval $[\underline{r},\overline{r}]$ (typically using some empirical $i$th quantile value of the sample as a bound).

The methods commonly used at present can be divided into two groups. A straightforward approach is to multiply the threshold value $r$ by the maximum $\ell(\bm{\theta})$  (see \cite{Wang2014}), i.e.
\begin{equation}\label{minr_cml}
\hat{r}_{{CML}}=\arg \max_{r \in [\underline{r},\overline{r}]}\ell(\bm{\theta}).
\end{equation}
Although this technique performs the best, the computational time is not ideal, especially for large sample size data since the CML score function's optimization speed is slow.

Another method of choosing threshold value is the minimum CLS score function $Q(\bm{\theta})$, see \cite{Moller2016}, \cite{LiY2018},
i.e.
\begin{equation*}
\hat{r}_{CLS}^{(1)}=\arg \min_{r \in [\underline{r},\overline{r}]}Q(\bm{\theta}).
\end{equation*}
Considering the computational speed and calculation burden, some researchers have provided algorithms for optimizing $\hat{r}_{CLS}^{(1)}$.
For example, \cite{LiY2018} proposed the doubly NeSS (D-NeSS) algorithm for the RCTINAR(1) process in view of \cite{LiT2016}'s nested subsample search (Ness) algorithm in the TAR model.
The D-Ness algorithm is also applicable to the model we proposed, and the details of the D-Ness algorithm are provided in the Appendix.

It should be mentioned that the precondition for using the D-Ness algorithm is that the conditional expectations of two regimes are distinct, so the performance of  $Q(\bm{\theta})$ is different for different threshold values $r$, i.e., $\phi_1\neq\phi_2$ is needed. However, $\phi_1\neq\phi_2$ is not necessary for the BiNB-MTTINAR(1) model. Considering this special case, we can choose the threshold $r$ in the following way. It is worth noting that the conditional variance of the two regimes differs, i.e., there is no $\phi_1,\phi_2,\lambda$ satisfying the following equation:
\begin{align*}
\phi_1(1-\phi_1)X_{t-1}+\lambda=\phi_2(1+\phi_2)X_{t-1}+\lambda(1+\lambda),
\end{align*}
for all $X_{t-1}\in \mathbb{N}_0$. Therefore, we can use the conditional variance as a criterion to select the threshold $r$. In view of the approach in \cite{Kar1988} and \cite{Ris2013}, we choose the threshold $r$ by minimizing the following score function:
\begin{equation}\label{Qn1}
\tilde{Q}_n(\bm{\theta},r)=\sum_{t=1}^n \left(V_t-[\phi_1(1-\phi_1)X_{t-1}+\lambda]I_{1,t}^R-[\phi_2(1+\phi_2)X_{t-1}+\lambda(1+\lambda)]I_{2,t}^R\right)^2,
\end{equation}
where $V_t=\left(X_t-\phi_1 X_{t-1}I_{1,t}^R-\phi_2 X_{t-1}I_{2,t}^R-\lambda\right)^2$. Although this method can also get an estimator of the parameter $\bm{\theta}$, it does not perform as well as CLS estimator $\bm{\hat{\theta}}_{CLS}$ (\ref{CLS_eq}). So the estimation of $\theta$ and $r$ can be done in the following two steps:\\
Step1. For each $r\in[\underline{r},\overline{r}]$, find $\hat{r}_{CLS}^{(2)}$ such that
\begin{align}
\hat{r}_{{CLS}}^{(2)}=\arg \min_{r \in [\underline{r},\overline{r}]}\tilde{Q}_n(\bm{\theta},r).
\end{align}
Step2. $\bm{\hat{\theta}}_{CLS}$ is estimated by (\ref{CLS_eq}) under $r=\hat{r}_{{CLS}}^{(2)}$.
\subsection{Wald Test}
Threshold models are typically characterized by piecewise linearization by partitioning a complex system into regimes by some threshold. Therefore, a hypothesis test for detecting the existence of the piecewise structure is highly desirable.
To date, many researchers have proposed different test statistics. A common and high-performance approach is to construct a likelihood ratio (LR) test based on the conditional likelihood function; see \cite{Moller2016}.
However, since the BiNB-MTTINAR(1) model is constructed by two operators, the LR test commonly used in TAR models cannot be effectively used to test the existence of a piecewise structure. Therefore, in our paper, we construct two Wald test statistics, also commonly used in threshold models, to detect the existence of the piecewise structure of the BiNB-MTTINAR(1) model.
\subsubsection{Wald Test for Conditional Expectation parameters}\label{sect3.4.1}
The null hypothesis and the alternative hypothesis take the form:
$$H_0^{(1)}:~\phi_1=\phi_2~~~~\text{vs.}~~~~H_1^{(1)}:~\phi_1\neq\phi_2.$$
That is, the existence of a piecewise structure is determined by checking whether the conditional expectation parameters are equal in two segment. A simple idea, learning from \cite{Yangtest2018}, is to use the test of the difference between two normal population means based on the asymptotic normality of some consistent estimators. Then we construct the Wald test based on the asymptotically distribution (\ref{clsnormal_eq}) of the CLS estimator $\hat{\bm{\theta}}_{CLS}$ and obtain the following result.
That is
\begin{align*}
T_{Wald-E}=\frac{n(\hat{\phi}_1-\hat{\phi_2})^2}{{\bm \Lambda}_{11}+{\bm \Lambda}_{22}-{\bm \Lambda}_{12}-{\bm \Lambda}_{21}},
\end{align*}
where ${\bm \Lambda}={\bm{V}}^{-1}{\bm{W}}{\bm{V}}^{-1}$ and $V, W$ are defined in Theorem \ref{clsnormal}.
In fact, according to the ergodicity of the BiNB-MTTINAR(1) model, it is easy to see that
$\widehat{{\bm \Lambda}}=\hat{\bm{V}}^{-1}\hat{\bm{W}}\hat{\bm{V}}^{-1}$, where $\hat{\bm{V}}$ and $\hat{\bm{W}}$ given by
\begin{align*}
\hat{\bm{V}}=\left( {\begin{array}{ccc}
\frac{1}{n}\sum_{t=1}^nX_{t-1}^2I_{1,t}^R&0&\frac{1}{n}\sum_{t=1}^nX_{t-1}I_{1,t}^R\\
0&\frac{1}{n}\sum_{t=1}^nX_{t-1}^2I_{2,t}^R&\frac{1}{n}\sum_{t=1}^nX_{t-1}I_{2,t}^R\\
\frac{1}{n}\sum_{t=1}^nX_{t-1}I_{1,t}^R&\frac{1}{n}\sum_{t=1}^nX_{t-1}I_{2,t}^R&1
\end{array}}\right),
\end{align*}
\begin{align*}
&\hat{\bm{W}}=\left( {\begin{array}{ccc}
\frac{1}{n}\sum_{t=1}^nU_{t}^2(\hat{\bm{\theta}})X_{t-1}^2I_{1,t}^R&0&\frac{1}{n}\sum_{t=1}^nU_{t}^2(\hat{\bm{\theta}})X_{t-1}I_{1,t}^R\\
0&\frac{1}{n}\sum_{t=1}^nU_{t}^2(\hat{\bm{\theta}})X_{t-1}^2I_{2,t}^R&\frac{1}{n}\sum_{t=1}^nU_{t}^2(\hat{\bm{\theta}})X_{t-1}I_{2,t}^R\\
\frac{1}{n}\sum_{t=1}^nU_{t}^2(\hat{\bm{\theta}})X_{t-1}I_{1,t}^R&\frac{1}{n}\sum_{t=1}^nU_{t}^2(\hat{\bm{\theta}})X_{t-1}I_{2,t}^R&\frac{1}{n}\sum_{t=1}^nU_{t}^2
\end{array}}\right),\\
&U_{t}(\hat{\bm{\theta}})=X_t-\hat{\phi}_{1,CLS} X_{t-1}I_{1,t}^R-\hat{\phi}_{2,CLS} X_{t-1}I_{2,t}^R-\hat{\lambda}_{CLS},
\end{align*}
are the consistent estimators of $\bm{V}$ and $\bm{W}$ (defined in Theorem \ref{clsnormal}).
Thus
\begin{align}\label{wald-e}
T_{Wald-E}=\frac{n(\hat{\phi}_1-\hat{\phi_2})^2}{\widehat{{\bm \Lambda}}_{11}+\widehat{{\bm \Lambda}}_{22}-\widehat{{\bm \Lambda}}_{12}-\widehat{{\bm \Lambda}}_{21}},
\end{align}
then under $H_0^{(1)}$,
\begin{align*}
T_{Wald-E}\overset{d}{\longrightarrow} \chi_1^2.
\end{align*}
\subsubsection{Wald Test for Conditional Variance parameters}
As discussed in Sect. 3.3, checking $\phi_1=\phi_2$ is not a complete indication that segmentation does not exist.
After all, because the operators in the two segments are different, even the $\phi_1=\phi_2$ segment structure still exists.
This leads to a problem: $T_{Wald-E}$ rejecting the null hypothesis can be used to justify that the piecewise structure exists, whereas $T_{Wald-E}$ accepting the null hypothesis cannot be used to support the conclusion that the segmented structure does not exist.
Considering this problem, we set the following null hypothesis and alternative hypothesis,
$$H_0^{(2)}:~\sigma_1^2=\sigma_2^2~~\text{and}~~b_1=b_2~~~~\text{vs.}~~~~H_1^{(2)}:~\sigma_1^2\neq\sigma_2^2~~\text{or}~~b_1\neq b_2.$$
where $\sigma_1^2={\rm Var}(B_i)$, $\sigma_2^2={\rm Var}(W_i)$, $b_1={\rm Var}(Z_{1,t})$, and $b_2={\rm Var}(Z_{2,t})$. $B_i$ and $W_i$ are defined in Definition \ref{BiNB}-(ii) and (iii).
In order to construct test statistics, we first derive the CLS estimators and asymptotic distributions of the conditional variance.

Let $\bm{\vartheta}=(\sigma_1^2,\sigma_2^2,b_1,b_2)^\textsf{T}$ be the conditional variance parameter vector, $$g_1(\bm{\vartheta},X_{t-1})={\rm Var}(X_t|X_{t-1})=(\sigma^2_1 X_{t-1}+b_1)I_{1,t}^R+(\sigma_2^2 X_{t-1}+b_2)I_{2,t}^R,$$
then the CLS estimator $\hat{\bm{\vartheta}}=(\hat{\sigma}_{1}^2,\hat{\sigma}_{2}^2,\hat{b}_{1},\hat{b}_{2})$ of $\bm{\vartheta}$ is obtained by minimizing the sum of the squared deviations,
\begin{align*}
S_n(\bm{\vartheta})&=\sum_{t=1}^nD_t^2(\bm{\vartheta})=\sum_{t=1}^n(V_t-g_1(\bm{\vartheta},X_{t-1}))^2 \\
&=\sum_{t=1}^n \left(V_t-(\sigma^2_1 X_{t-1}+b_1)I_{1,t}^R-(\sigma_2^2 X_{t-1}+b_2)I_{2,t}^R\right)^2,\nonumber
\end{align*}
where $V_t=\left(X_t-\hat{\phi}_{1,CLS} X_{t-1}I_{1,t}^R-\hat{\phi}_{2,CLS} X_{t-1}I_{2,t}^R-\hat{\lambda}_{CLS}\right)^2$,
$\hat{\phi}_{1,CLS}$, $\hat{\phi}_{2,CLS}$ and $\hat{\lambda}_{CLS}$ are the CLS estimators from (\ref{CLS_eq}). Furthermore, $\hat{\bm{\vartheta}}$ is a consistent and asymptotically normal estimator of $\bm{\vartheta}$.
\begin{thm}\label{consist-var}
Let $\{X_t\}$ be a BiNB-MTTINAR(1) process, then the CLS estimator $\hat{\bm{\vartheta}}$ is consistent and has the following asymptotically distribution,
\begin{align}\label{cls_var}
\sqrt{n}(\hat{\bm{\vartheta}}-\bm{\vartheta}_0) \overset{d}{\longrightarrow}N(\bm{0},\bm{\Sigma}),
\end{align}
where $\bm{\Sigma}=\bm{\tilde{V}}^{-1}\bm{\tilde{W}}\bm{\tilde{V}}^{-1}$, $\bm{\tilde{V}}$ and $\bm{\tilde{W}}$ are square matrices of order 4 with the $ij$th element given by
$$\tilde{V}_{ij}:={\rm E}\left(\frac{\partial}{\partial\bm{\vartheta}_{i}}g_1(\bm{\vartheta},X_{t-1})\frac{\partial}{\partial\bm{\vartheta}_{j}}g_1(\bm{\vartheta},X_{t-1})\right)_{\bm{\vartheta}_0},$$
$$\tilde{W}_{ij}:={\rm E}\left(D_{t}^2(\bm{\vartheta})\frac{\partial}{\partial\bm{\vartheta}_i}g_1(\bm{\vartheta},X_{t-1})\frac{\partial}{\partial\bm{\vartheta}_j}g_1(\bm{\vartheta},X_{t-1})\right)_{\bm{\vartheta}_0}.$$
\end{thm}
Furthermore, Denote $\hat{\bm{\Sigma}}=\hat{\tilde{\bm{V}}}^{-1}\hat{\tilde{\bm{W}}}\hat{\tilde{\bm{V}}}^{-1}$ and
\begin{align*}
\hat{\tilde{\bm{V}}}=\left( {\begin{array}{cccc}
\frac{1}{n}\sum_{t=1}^nX_{t-1}^2I_{1,t}^R&0&\frac{1}{n}\sum_{t=1}^nX_{t-1}I_{1,t}^R&0\\
0&\frac{1}{n}\sum_{t=1}^nX_{t-1}^2I_{2,t}^R&0&\frac{1}{n}\sum_{t=1}^nX_{t-1}I_{2,t}^R\\
\frac{1}{n}\sum_{t=1}^nX_{t-1}I_{1,t}^R&0&\frac{1}{n}\sum_{t=1}^nI_{1,t}^R&0\\
0&\frac{1}{n}\sum_{t=1}^nX_{t-1}I_{2,t}^R&0&\frac{1}{n}\sum_{t=1}^nI_{1,t}^R
\end{array}}\right),
\end{align*}
\begin{align*}
&\hat{\tilde{\bm{W}}}=\left( {\begin{array}{cccc}
\frac{1}{n}\sum_{t=1}^nD_{t}^2(\hat{\bm{\vartheta}})X_{t-1}^2I_{1,t}^R&0&\frac{1}{n}\sum_{t=1}^nD_{t}^2(\hat{\bm{\vartheta}})X_{t-1}I_{1,t}^R&0\\
0&\frac{1}{n}\sum_{t=1}^nD_{t}^2(\hat{\bm{\vartheta}})X_{t-1}^2I_{2,t}^R&0&\frac{1}{n}\sum_{t=1}^nD_{t}^2(\hat{\bm{\vartheta}})X_{t-1}I_{2,t}^R\\
\frac{1}{n}\sum_{t=1}^nD_{t}^2(\hat{\bm{\vartheta}})X_{t-1}I_{1,t}^R&0&\frac{1}{n}\sum_{t=1}^nD_{t}^2(\hat{\bm{\vartheta}})I_{1,t}^R&0\\
0&\frac{1}{n}\sum_{t=1}^nD_{t}^2(\hat{\bm{\vartheta}})X_{t-1}I_{2,t}^R&0&\frac{1}{n}\sum_{t=1}^nD_{t}^2(\hat{\bm{\vartheta}})I_{1,t}^R
\end{array}}\right),\\
&D_{t}(\hat{\bm{\vartheta}})=V_t-(\hat{\sigma}^2_1 X_{t-1}+\hat{b}_1)I_{1,t}^R-(\hat{\sigma}_2^2 X_{t-1}+\hat{b}_2)I_{2,t}^R,\\
&V_t=(X_t-\hat{\phi}_{1,CLS} X_{t-1}I_{1,t}^R-\hat{\phi}_{2,CLS} X_{t-1}I_{2,t}^R-\hat{\lambda}_{CLS})^2.
\end{align*}
Clearly, $\hat{\bm{\Sigma}}$ is the consistent estimator of $\bm{\Sigma}$.
Then similar to the test statistics constructed in Sect. \ref{sect3.4.1}, we also use a test for the difference between two normal population means based on the asymptotic normality of the conditional variance parameter estimators. That is we construct the Wald test based on the asymptotically distribution (\ref{cls_var}) of the CLS estimator $\bm{\hat{\vartheta}}$ and obtain the following result.
Define
\begin{align}\label{wald-var}
T_{Wald-Var}=\frac{n(\hat{\sigma}_1^2-\hat{\sigma}_2^2)^2}{\hat{\bm{\Sigma}}_{11}+\hat{\bm{\Sigma}}_{22}-\hat{\bm{\Sigma}}_{12}-\hat{\bm{\Sigma}}_{21}}
+\frac{n(\hat{b}_1-\hat{b}_2)^2}{\hat{\bm{\Sigma}}_{33}+\hat{\bm{\Sigma}}_{44}-\hat{\bm{\Sigma}}_{34}-\hat{\bm{\Sigma}}_{43}},
\end{align}
Then under $H_0^{(2)}$,
\begin{align*}
T_{Wald-Var}\overset{d}{\longrightarrow} \chi_2^2.
\end{align*}
\section{Simulation Studies}
In this section, we first compare the performance of the proposed estimators under the case that $r$ is known. Then, we investigate the performance of the threshold estimation methods proposed in Sect. 3.2. 
Finally, the empirical size and power of the Wald test are explored in Sect. 4.3.

Let us start by introducing the simulation study settings. The dataset $\{X_t\}_{t=1}^n$ is generated from the BiNB-MTTINAR(1) model (\ref{Mymodel}): A1 - A4 (B1 - B4). The parameter settings are shown in Table \ref{tab1}.
For each model, the value of $r$ is chosen such that the observations in each regime comprise at least 20\% of the total sample size. All simulations are conducted using MATLAB. The empirical results displayed in the tables and box plots, that is, the empirical biases and mean square errors (MSE), are computed over $10000$ replications.
\begin{table}[H]
\renewcommand\arraystretch{0.9}      
\caption{Parameters setting of different models}
\label{tab1}
\vspace{-3mm}                    
\centering                            
{\tabcolsep0.15in                     
\begin{tabular}{ccccccccccc}
  \hline
  &  \multicolumn{4}{c}{$R=0$} & & & \multicolumn{4}{c}{$R=1$}\\
  \cline{2-5}\cline{8-11}
Model& $\phi_1$ & $\phi_2$ & $\lambda$ & $r$ & &Model&  $\phi_1$ & $\phi_2$ & $\lambda$ & $r$ \\
\hline
 A1  & 0.4 &   0.2  & 3&  4   & & B1 & 0.4  &  0.2  & 3 &  4  \\
  A2  & 0.4 &   0.4  & 3&  4   & & B2 & 0.4  &  0.4  & 3 &  4  \\
 A3  & 0.3 & 0.6  &  5  &  7  & & B3 & 0.3  &  0.6 & 5 & 7  \\
 A4 & 0.6 &  0.6 & 5  &  12  & & B4 & 0.6  &  0.6 & 5 &  12 \\
\hline
\end{tabular}
}
\end{table}

\begin{figure}
\begin{adjustwidth}{-1.5cm}{-1cm}
\includegraphics[width=7.4in,height=6.5in]{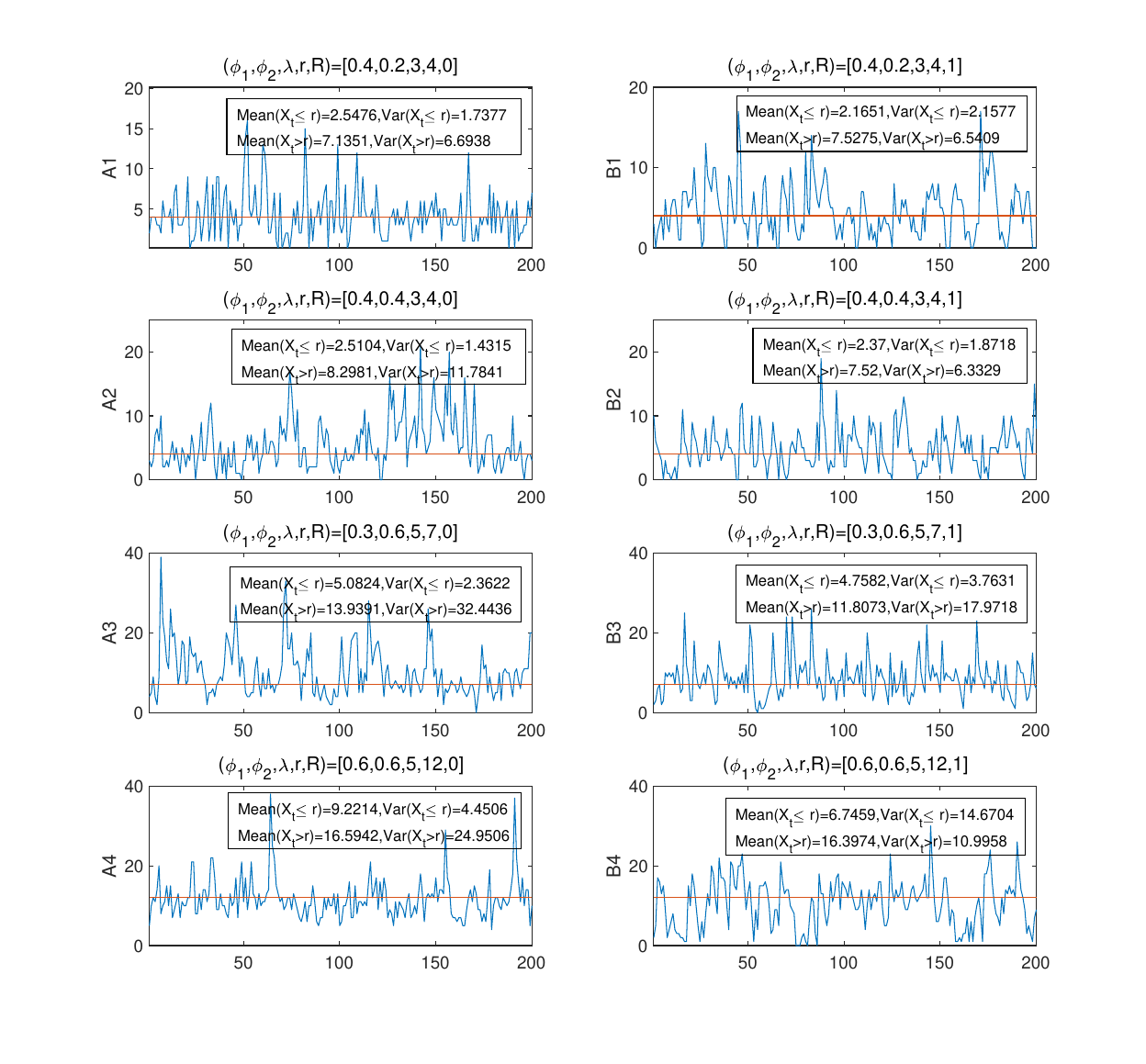}
\end{adjustwidth}
\vspace{-16mm}
\caption{Sample paths plots of models A1 - A4 and B1 - B4. \protect\\The red lines in the sample paths are the threshold values of each models.}
\label{pathA}
\end{figure}

Fig.\ref{pathA} shows the sample paths for Models A1 - B4 of the BiNB-MTTINAR(1) process and the mean and variance of the two regimes. Comparing the sample paths shown in Fig.\ref{pathA}, we find that under the same parameter values, it is easier to distinguish the two regimes in the A1 - A4 models. This implies that when $R=0$, distinguishing between the two regimes is easy. Meanwhile, by comparing the variance difference of the two regimes in the left and right halves of the figure, it can also be seen that, as discussed in Remark in Sect. 2, the variance difference between the two regimes of the BiNB-MTTINAR(1) model is more obvious when $R=0$, especially when $\phi_2$ is not less than $\phi_1$.
\subsection{Simulation study when $r$ is known}
\begin{landscape}
\begin{figure}[H]
\begin{adjustwidth}{-1.5cm}{-1cm}
 \includegraphics[width=11in,height=5.5in]{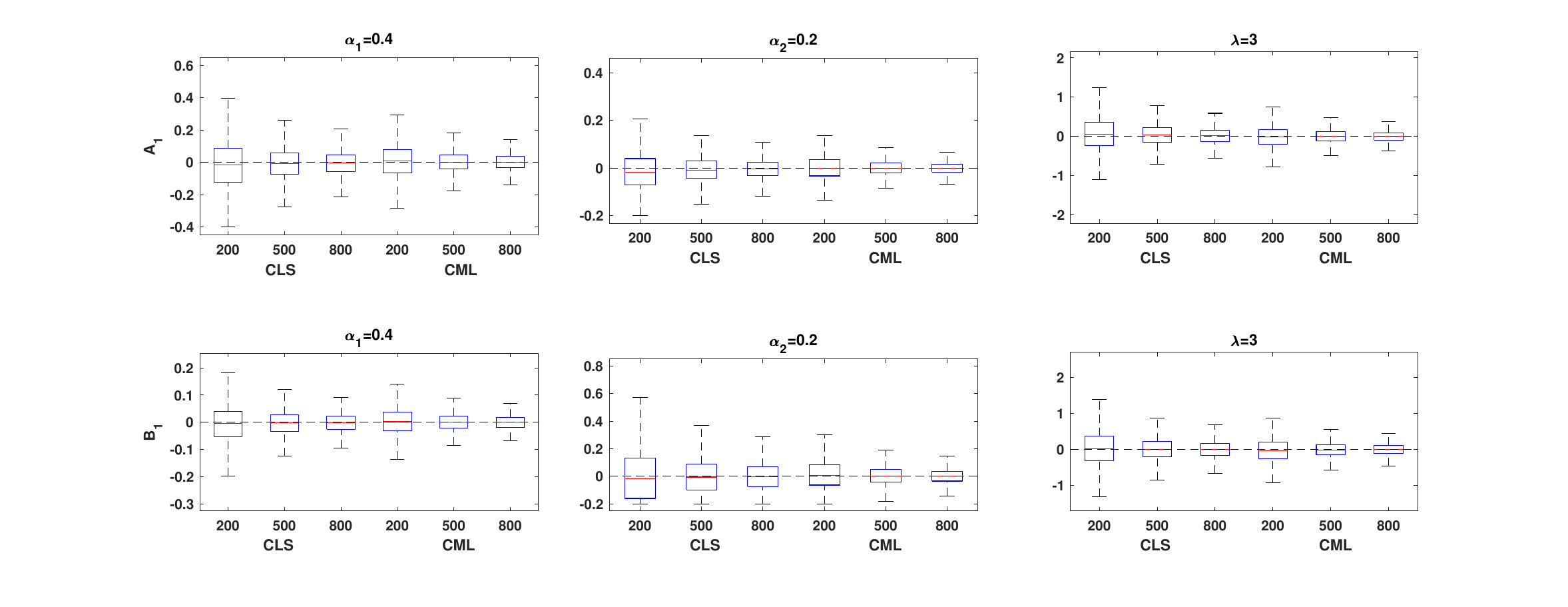}
 \end{adjustwidth}
\vspace{-12mm}
\caption{Box plots from 10000 CLS and CML simulation estimators for models A1 and B1, sample size $n=200, 500, 800$.}\label{box_r_known}
\end{figure}
\end{landscape}

\begin{figure}[H]
\begin{adjustwidth}{-2cm}{-2cm}
 \includegraphics[width=8in,height=7in]{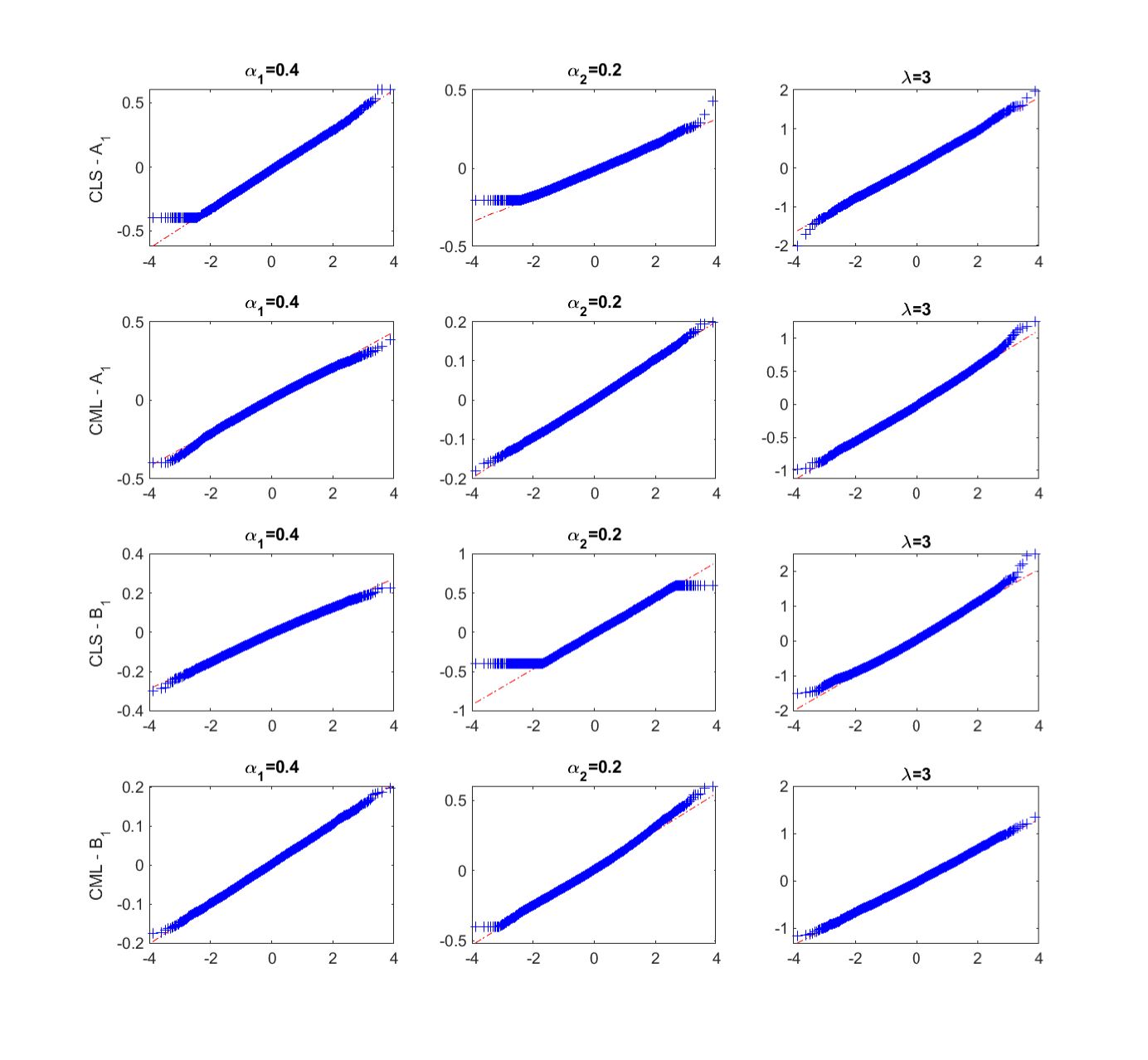}
\end{adjustwidth}
\vspace{-12mm}
\caption{QQ plots of CLS and CML estimators for models A1 and B1, sample size $n=200$.}\label{qq-plot}
\end{figure}

Table \ref{r-known1} and Table \ref{r-known2} report the bias and MSE of the CLS and CML estimators for Models A1 - B4 when $r$ is known. The sample sizes considered are $n = 200, 500$ and $800$. From these two tables, it is easy to see that all the simulation results perform better as $n$ increases, which implies that the two estimation methods can lead to good and consistent estimators when $r$ is known.
Moreover, $\hat{\bm{\theta}}_{CML}$ has a smaller Bias and MSE, which implies that $\hat{\bm{\theta}}_{CML}$ performs better than $\hat{\bm{\theta}}_{CLS}$.
\begin{table}
\scriptsize
\begin{adjustwidth}{0.5cm}{1cm}
\caption{Simulation results for models A1 - B4 when $r$ is known}\label{r-known1} 
\vspace{-3mm}   
\renewcommand\arraystretch{0.8}        
{\tabcolsep0.1in                     
\begin{tabular}{cc*{11}{r}}
\hline
&&\multicolumn{5}{c}{A1}&&\multicolumn{5}{c}{A2}\\\cline{3-7}\cline{9-13}
&&\multicolumn{2}{c}{CLS}&&\multicolumn{2}{c}{CML}&&\multicolumn{2}{c}{CLS}&&\multicolumn{2}{c}{CML}\\\cline{3-4}\cline{6-7}\cline{9-10}\cline{12-13}
$n$&Para.&\multicolumn{1}{c}{Bias}&\multicolumn{1}{c}{MSE}&&\multicolumn{1}{c}{Bias}&\multicolumn{1}{c}{MSE}&&\multicolumn{1}{c}{Bias}&\multicolumn{1}{c}{MSE}&&\multicolumn{1}{c}{Bias}&\multicolumn{1}{c}{MSE}\\\hline
200&$\phi_1$&$-$0.0176 &0.0235 &&0.0047 &0.0113 &&$-$0.0269 &0.0354 &&0.0023 &0.0139 \\
&$\phi_2$&$-$0.0153 &0.0070 &&0.0013 &0.0025 &&$-$0.0199 &0.0080 &&$-$0.0019 &0.0029 \\
&$\lambda$&0.0608 &0.1932 &&$-$0.0125 &0.0811 &&0.1050 &0.3347 &&0.0007 &0.1067 \\
500&$\phi_1$&$-$0.0090 &0.0100 &&0.0017 &0.0045 &&$-$0.0146 &0.0149 &&0.0013 &0.0053 \\
&$\phi_2$&$-$0.0069 &0.0029 &&0.0007 &0.0010 &&$-$0.0091 &0.0032 &&$-$0.0009 &0.0012 \\
&$\lambda$&0.0310 &0.0808 &&$-$0.0031 &0.0323 &&0.0486 &0.1346 &&$-$0.0027 &0.0419 \\
800&$\phi_1$&$-$0.0043 &0.0063 &&0.0007 &0.0028 &&$-$0.0074 &0.0095 &&0.0004 &0.0033 \\
&$\phi_2$&$-$0.0037 &0.0018 &&0.0001 &0.0006 &&$-$0.0053 &0.0020 &&$-$0.0008 &0.0007 \\
&$\lambda$&0.0138 &0.0499 &&$-$0.0026 &0.0191 &&0.0263 &0.0856 &&0.0006 &0.0263 \\
\hline
&&\multicolumn{5}{c}{A3}&&\multicolumn{5}{c}{A4}\\\cline{3-7}\cline{9-13}
&&\multicolumn{2}{c}{CLS}&&\multicolumn{2}{c}{CML}&&\multicolumn{2}{c}{CLS}&&\multicolumn{2}{c}{CML}\\\cline{3-4}\cline{6-7}\cline{9-10}\cline{12-13}
$n$&Para.&\multicolumn{1}{c}{Bias}&\multicolumn{1}{c}{MSE}&&\multicolumn{1}{c}{Bias}&\multicolumn{1}{c}{MSE}&&\multicolumn{1}{c}{Bias}&\multicolumn{1}{c}{MSE}&&\multicolumn{1}{c}{Bias}&\multicolumn{1}{c}{MSE}\\\hline
200&$\phi_1$&$-$0.0298 &0.0348 &&0.0040 &0.0133 &&$-$0.0175 &0.0141 &&0.0037 &0.0036 \\
&$\phi_2$&$-$0.0229 &0.0068 &&$-$0.0031 &0.0019 &&$-$0.0160 &0.0066 &&0.0016 &0.0017 \\
&$\lambda$&0.2493 &1.2945 &&$-$0.0072 &0.3385 &&0.1783 &1.1695 &&$-$0.0262 &0.2748 \\
500&$\phi_1$&$-$0.0162 &0.0169 &&0.0016 &0.0052 &&$-$0.0073 &0.0057 &&0.0017 &0.0014 \\
&$\phi_2$&$-$0.0091 &0.0026 &&$-$0.0010 &0.0007 &&$-$0.0073 &0.0027 &&0.0001 &0.0007 \\
&$\lambda$&0.1031 &0.5114 &&$-$0.0032 &0.1308 &&0.0740 &0.4771 &&$-$0.0125 &0.1094 \\
800&$\phi_1$&$-$0.0121 &0.0110 &&0.0006 &0.0033 &&$-$0.0043 &0.0036 &&0.0012 &0.0009 \\
&$\phi_2$&$-$0.0064 &0.0016 &&$-$0.0008 &0.0005 &&$-$0.0043 &0.0017 &&0.0004 &0.0004 \\
&$\lambda$&0.0715 &0.3241 &&$-$0.0018 &0.0823 &&0.0443 &0.2989 &&$-$0.0091 &0.0680 \\
\hline
\end{tabular}
}
\end{adjustwidth}
\end{table}
\begin{table}
\scriptsize
\begin{adjustwidth}{0.5cm}{1cm}
\caption{Simulation results for models B1 - B4 when $r$ is known}
\label{r-known2}                    
\renewcommand\arraystretch{0.8}        
\vspace{-3mm}                       
{\tabcolsep0.1in                     
\begin{tabular}{cc*{11}{r}}
\hline
&&\multicolumn{5}{c}{B1}&&\multicolumn{5}{c}{B2}\\\cline{3-7}\cline{9-13}
&&\multicolumn{2}{c}{CLS}&&\multicolumn{2}{c}{CML}&&\multicolumn{2}{c}{CLS}&&\multicolumn{2}{c}{CML}\\\cline{3-4}\cline{6-7}\cline{9-10}\cline{12-13}
$n$&Para.&\multicolumn{1}{c}{Bias}&\multicolumn{1}{c}{MSE}&&\multicolumn{1}{c}{Bias}&\multicolumn{1}{c}{MSE}&&\multicolumn{1}{c}{Bias}&\multicolumn{1}{c}{MSE}&&\multicolumn{1}{c}{Bias}&\multicolumn{1}{c}{MSE}\\\hline
200&$\phi_1$&$-$0.0071 &0.0050 &&0.0023 &0.0026 &&$-$0.0075 &0.0056 &&0.0029 &0.0028 \\
&$\phi_2$&0.0104 &0.0344 &&0.0140 &0.0127 &&$-$0.0091 &0.0477 &&0.0131 &0.0193 \\
&$\lambda$&0.0434 &0.2559 &&$-$0.0250 &0.1118 &&0.0491 &0.3044 &&$-$0.0257 &0.1345 \\
500&$\phi_1$&$-$0.0024 &0.0020 &&0.0009 &0.0010 &&$-$0.0029 &0.0022 &&0.0011 &0.0011 \\
&$\phi_2$&0.0002 &0.0167 &&0.0047 &0.0048 &&$-$0.0034 &0.0209 &&0.0068 &0.0073 \\
&$\lambda$&0.0156 &0.1032 &&$-$0.0090 &0.0454 &&0.0169 &0.1214 &&$-$0.0116 &0.0528 \\
800&$\phi_1$&$-$0.0018 &0.0012 &&$-$0.0001 &0.0007 &&$-$0.0022 &0.0014 &&0.0002 &0.0007 \\
&$\phi_2$&$-$0.0004 &0.0112 &&0.0022 &0.0029 &&$-$0.0015 &0.0128 &&0.0040 &0.0045 \\
&$\lambda$&0.0102 &0.0632 &&$-$0.0018 &0.0282 &&0.0129 &0.0725 &&$-$0.0038 &0.0321 \\
\hline
&&\multicolumn{5}{c}{B3}&&\multicolumn{5}{c}{B4}\\\cline{3-7}\cline{9-13}
&&\multicolumn{2}{c}{CLS}&&\multicolumn{2}{c}{CML}&&\multicolumn{2}{c}{CLS}&&\multicolumn{2}{c}{CML}\\\cline{3-4}\cline{6-7}\cline{9-10}\cline{12-13}
$n$&Para.&\multicolumn{1}{c}{Bias}&\multicolumn{1}{c}{MSE}&&\multicolumn{1}{c}{Bias}&\multicolumn{1}{c}{MSE}&&\multicolumn{1}{c}{Bias}&\multicolumn{1}{c}{MSE}&&\multicolumn{1}{c}{Bias}&\multicolumn{1}{c}{MSE}\\\hline
200&$\phi_1$&$-$0.0046 &0.0057 &&0.0055 &0.0025 &&$-$0.0117 &0.0046 &&0.0002 &0.0013 \\
&$\phi_2$&$-$0.0083 &0.0422 &&0.0134 &0.0129 &&$-$0.0129 &0.0283 &&0.0075 &0.0057 \\
&$\lambda$&0.0518 &0.8338 &&$-$0.0637 &0.3060 &&0.1925 &1.4799 &&$-$0.0204 &0.3804 \\
500&$\phi_1$&$-$0.0022 &0.0023 &&0.0020 &0.0010 &&$-$0.0044 &0.0017 &&$-$0.0002 &0.0005 \\
&$\phi_2$&$-$0.0030 &0.0173 &&0.0058 &0.0049 &&$-$0.0058 &0.0113 &&0.0021 &0.0021 \\
&$\lambda$&0.0226 &0.3347 &&$-$0.0261 &0.1206 &&0.0712 &0.5467 &&$-$0.0045 &0.1529 \\
800&$\phi_1$&$-$0.0010 &0.0014 &&0.0015 &0.0006 &&$-$0.0024 &0.0010 &&0.0000 &0.0003 \\
&$\phi_2$&$-$0.0016 &0.0104 &&0.0035 &0.0030 &&$-$0.0028 &0.0073 &&0.0018 &0.0013 \\
&$\lambda$&0.0121 &0.2022 &&$-$0.0170 &0.0734 &&0.0371 &0.3412 &&$-$0.0067 &0.0950 \\
\hline
\end{tabular}
}
\end{adjustwidth}
\end{table}
Fig.\ref{box_r_known} is obtained from the bias of 10000 CLS and CML simulation estimators for Models A1 and B1. Note that the box plots are symmetric and centered on zero-bias, both bias and MSE for the CML estimators are smaller than the CLS estimators, which confirms the previous conclusions. Fig.\ref{qq-plot} shows the QQ plots of the CLS and CML estimators for Models A1 and B1 with sample size $n=200$. From the QQ plots, we can see that the CLS and CML estimators are asymptotically normal for all parameters. Similar results are obtained for the remaining models, and the figures are omitted here to save space.
\subsection{Simulation study under $r$ is unknown}
Table \ref{r-unknown1} and Table \ref{r-unknown2} report the bias and MSE of the CLS and CML estimators proposed in Set. 3.3 for Models A1 - B4. The considered sample sizes are $n = 200, 500, 800$ and $1500$. The search range of the threshold (i.e., $[\underline{r},\overline{r}]$) is set to the empirical 10th and 90th quantile of $\{X_t\}_{t=1}^n$.
From these two tables, it is clear that both the bias and MSE of the CML estimators are very small, as expected. For the CLS estimators, where $r$ is estimated by $r_{CLS}^{(2)}$, although their bias and MSE are larger than those of the CML estimators, they both decrease with the increase of $n$. That is, both methods are reliable, which ensures that we can confidently use the BiNB-MTTINAR(1) model without worrying about incorrect results when $r$ is unknown. A detailed explanation and analysis of threshold estimators are presented in Table \ref{r-estimate}.
\begin{table}[H]             
\scriptsize
\begin{adjustwidth}{0.5cm}{1cm}
\caption{Simulation results for models A1 - A4 when $r$ is unknown}
\label{r-unknown1}                      
\renewcommand\arraystretch{0.8}        
\vspace{-3mm}                       
{\tabcolsep0.1in                     
\begin{tabular}{cc*{11}{r}}
\hline
&&\multicolumn{5}{c}{A1}&&\multicolumn{5}{c}{A2}\\\cline{3-7}\cline{9-13}
&&\multicolumn{2}{c}{CLS}&&\multicolumn{2}{c}{CML}&&\multicolumn{2}{c}{CLS}&&\multicolumn{2}{c}{CML}\\\cline{3-4}\cline{6-7}\cline{9-10}\cline{12-13}
$n$&Para.&Bias&MSE&&Bias&MSE&&Bias&MSE&&Bias&MSE\\\hline
200&$\phi_1$&$-$0.1195 &0.0486 &&0.0042 &0.0116 &&$-$0.0423 &0.0327 &&$-$0.0005 &0.0154 \\
&$\phi_2$&$-$0.0268 &0.0088 &&0.0008 &0.0025 &&$-$0.0228 &0.0090 &&$-$0.0040 &0.0030 \\
&$\lambda$&0.2809 &0.3292 &&$-$0.0096 &0.0820 &&0.1488 &0.3160 &&0.0088 &0.1142 \\
&$r$&1.1117 &4.1479 &&0.0133 &0.0329 &&1.6532 &8.5216 &&0.0399 &0.1349 \\
500&$\phi_1$&$-$0.0731 &0.0249 &&0.0019 &0.0044 &&$-$0.0257 &0.0141 &&0.0016 &0.0052 \\
&$\phi_2$&$-$0.0147 &0.0037 &&0.0012 &0.0010 &&$-$0.0090 &0.0034 &&$-$0.0008 &0.0011 \\
&$\lambda$&0.1597 &0.1437 &&$-$0.0052 &0.0320 &&0.0811 &0.1278 &&$-$0.0031 &0.0406 \\
&$r$&0.5716 &1.9082 &&0.0001 &0.0001 &&0.8042 &4.1302 &&0.0012 &0.0020 \\
800&$\phi_1$&$-$0.0479 &0.0156 &&0.0011 &0.0028 &&$-$0.0163 &0.0095 &&0.0008 &0.0033 \\
&$\phi_2$&$-$0.0105 &0.0022 &&0.0002 &0.0006 &&$-$0.0063 &0.0021 &&$-$0.0007 &0.0007 \\
&$\lambda$&0.1051 &0.0894 &&$-$0.0024 &0.0204 &&0.0511 &0.0850 &&$-$0.0018 &0.0252 \\
&$r$&0.3502 &1.1574 &&0 &0 &&0.4330 &2.1804 &&0.0001 &0.0001 \\
1500&$\phi_1$&$-$0.0208 &0.0071 &&0.0002 &0.0015 &&$-$0.0092 &0.0050 &&0.0003 &0.0017 \\
&$\phi_2$&$-$0.0049 &0.0011 &&0.0003 &0.0003 &&$-$0.0041 &0.0011 &&$-$0.0004 &0.0004 \\
&$\lambda$&0.0452 &0.0419 &&$-$0.0006 &0.0106 &&0.0301 &0.0454 &&0.0011 &0.0137 \\
&$r$&0.1370 &0.3988 &&0 &0 &&0.1195 &0.5153 &&0 &0 \\
\hline
&&\multicolumn{5}{c}{A3}&&\multicolumn{5}{c}{A4}\\\cline{3-7}\cline{9-13}
&&\multicolumn{2}{c}{CLS}&&\multicolumn{2}{c}{CML}&&\multicolumn{2}{c}{CLS}&&\multicolumn{2}{c}{CML}\\\cline{3-4}\cline{6-7}\cline{9-10}\cline{12-13}
$n$&Para.&Bias&MSE&&Bias&MSE&&Bias&MSE&&Bias&MSE\\\hline
200&$\phi_1$&0.0788&0.0649&&0.0048&0.0151&&$-$0.0630&0.0258&&0.0003&0.0040\\
&$\phi_2$&$-$0.0058&0.0063&&$-$0.0035&0.0019&&$-$0.0360&0.0084&&$-$0.0051&0.0021\\
&$\lambda$&$-$0.0675&1.5726&&0.0010&0.3626&&0.6224&2.2781&&0.0086&0.3199\\
&$r$&2.2616&20.6480&&0.0038&0.1084&&2.0816&15.9816&&0.0320&0.0970\\
500&$\phi_1$&0.0249&0.0262&&0.0020&0.0054&&$-$0.0344&0.0100&&0.0006&0.0014\\
&$\phi_2$&$-$0.0033&0.0027&&$-$0.0015&0.0007&&$-$0.0154&0.0032&&$-$0.0014&0.0008\\
&$\lambda$&$-$0.0021&0.5952&&$-$0.0040&0.1375&&0.3270&0.8893&&$-$0.0027&0.1181\\
&$r$&0.7692&6.6242&&$-$0.0010&0.0036&&1.0950&9.0246&&0.0008&0.0020\\
800&$\phi_1$&0.0063&0.0159&&0.0003&0.0033&&$-$0.0228&0.0064&&0.0006&0.0009\\
&$\phi_2$&$-$0.0038&0.0017&&$-$0.0007&0.0005&&$-$0.0105&0.0021&&$-$0.0006&0.0005\\
&$\lambda$&0.0362&0.3725&&0.0038&0.0848&&0.2177&0.5653&&$-$0.0006&0.0741\\
&$r$&0.3615&3.0955&&$-$0.0002&0.0002&&0.6005&4.7813&&0.0002&0.0002\\
1500&$\phi_1$&$-$0.0024&0.0069&&0&0.0017&&$-$0.0102&0.0032&&0.0001&0.0005\\
&$\phi_2$&$-$0.0027&0.0009&&$-$0.0004&0.0002&&$-$0.0055&0.0011&&$-$0.0006&0.0003\\
&$\lambda$&0.0317&0.1857&&0.0030&0.0444&&0.0956&0.2897&&$-$0.0007&0.0395\\
&$r$&0.0702&0.5684&&0&0&&0.2046&1.5148&&0&0\\
\hline
\end{tabular}
}
\end{adjustwidth}

\end{table}

\begin{table}[H]             
\scriptsize
\begin{adjustwidth}{0.5cm}{1cm}
\caption{Simulation results for models B1 - B4 when $r$ is unknown}
\label{r-unknown2}                      
\renewcommand\arraystretch{0.8}        
\vspace{-3mm}                       
{\tabcolsep0.1in                     
\begin{tabular}{cc*{11}{r}}
\hline
&&\multicolumn{5}{c}{B1}&&\multicolumn{5}{c}{B2}\\\cline{3-7}\cline{9-13}
&&\multicolumn{2}{c}{CLS}&&\multicolumn{2}{c}{CML}&&\multicolumn{2}{c}{CLS}&&\multicolumn{2}{c}{CML}\\\cline{3-4}\cline{6-7}\cline{9-10}\cline{12-13}
$n$&Para.&Bias&MSE&&Bias&MSE&&Bias&MSE&&Bias&MSE\\\hline
200&$\phi_1$&0.0266 &0.0053 &&0.0028 &0.0026 &&0.0125 &0.0053 &&0.0029 &0.0029 \\
&$\phi_2$&0.2709 &0.1771 &&0.0174 &0.0136 &&0.1612 &0.1103 &&0.0159 &0.0212 \\
&$\lambda$&$-$0.3076 &0.3341 &&$-$0.0281 &0.1159 &&$-$0.1382 &0.2822 &&$-$0.0312 &0.1412 \\
&$r$&$-$1.1210 &2.4938 &&0.0225 &0.0899 &&$-$0.9897 &2.1701 &&0.0146 &0.1232 \\
500&$\phi_1$&0.0185 &0.0023 &&0.0009 &0.0010 &&0.0056 &0.0021 &&0.0005 &0.0011 \\
&$\phi_2$&0.1733 &0.0944 &&0.0060 &0.0049 &&0.0806 &0.0524 &&0.0050 &0.0071 \\
&$\lambda$&$-$0.2134 &0.1603 &&$-$0.0079 &0.0454 &&$-$0.0615 &0.1106 &&$-$0.0061 &0.0522 \\
&$r$&$-$0.8806 &1.8220 &&0.0006 &0.0008 &&$-$0.7650 &1.6374 &&0 &0.0030 \\
800&$\phi_1$&0.0148 &0.0015 &&0.0007 &0.0006 &&0.0045 &0.0013 &&0.0004 &0.0007 \\
&$\phi_2$&0.1327 &0.0635 &&0.0040 &0.0030 &&0.0598 &0.0330 &&0.0038 &0.0045 \\
&$\lambda$&$-$0.1755 &0.1101 &&$-$0.0069 &0.0273 &&$-$0.0480 &0.0719 &&$-$0.0045 &0.0328 \\
&$r$&$-$0.7371 &1.4397 &&0 &0 &&$-$0.6154 &1.2624 &&0 &0 \\
1500&$\phi_1$&0.0105 &0.0009 &&0.0002 &0.0003 &&0.0026 &0.0007 &&0.0003 &0.0004 \\
&$\phi_2$&0.0883 &0.0343 &&0.0026 &0.0016 &&0.0305 &0.0150 &&0.0024 &0.0024 \\
&$\lambda$&$-$0.1277 &0.0680 &&$-$0.0035 &0.0148 &&$-$0.0274 &0.0373 &&$-$0.0038 &0.0174 \\
&$r$&$-$0.5483 &0.9755 &&0 &0 &&$-$0.4118 &0.7958 &&0 &0 \\
\hline
&&\multicolumn{5}{c}{B3}&&\multicolumn{5}{c}{B4}\\\cline{3-7}\cline{9-13}
&&\multicolumn{2}{c}{CLS}&&\multicolumn{2}{c}{CML}&&\multicolumn{2}{c}{CLS}&&\multicolumn{2}{c}{CML}\\\cline{3-4}\cline{6-7}\cline{9-10}\cline{12-13}
$n$&Para.&Bias&MSE&&Bias&MSE&&Bias&MSE&&Bias&MSE\\\hline
200&$\phi_1$&$-$0.0104 &0.0064 &&0.0048 &0.0026 &&0.0201 &0.0046 &&0.0013 &0.0014 \\
&$\phi_2$&0.0102 &0.0602 &&0.0131 &0.0132 &&0.1237 &0.0528 &&0.0062 &0.0065 \\
&$\lambda$&0.2001 &1.0179 &&$-$0.0587 &0.3162 &&$-$0.4410 &1.3909 &&$-$0.0414 &0.3708 \\
&$r$&$-$0.9741 &2.7825 &&$-$0.0062 &0.0780 &&$-$2.3478 &11.9148 &&$-$0.0061 &0.0979 \\
500&$\phi_1$&$-$0.0107 &0.0030 &&0.0019 &0.0010 &&0.0104 &0.0019 &&0.0004 &0.0006 \\
&$\phi_2$&$-$0.0238 &0.0320 &&0.0059 &0.0050 &&0.0713 &0.0239 &&0.0029 &0.0024 \\
&$\lambda$&0.1927 &0.5398 &&$-$0.0217 &0.1217 &&$-$0.2285 &0.5555 &&$-$0.0138 &0.1420 \\
&$r$&$-$0.6208 &1.7522 &&$-$0.0002 &0.0014 &&$-$1.8164 &8.4670 &&$-$0.0004 &0.0026 \\
800&$\phi_1$&$-$0.0082 &0.0020 &&0.0011 &0.0006 &&0.0072 &0.0011 &&0.0003 &0.0003 \\
&$\phi_2$&$-$0.0241 &0.0210 &&0.0028 &0.0030 &&0.0475 &0.0130 &&0.0021 &0.0015 \\
&$\lambda$&0.1456 &0.3621 &&$-$0.0152 &0.0742 &&$-$0.1610 &0.3399 &&$-$0.0113 &0.0867 \\
&$r$&$-$0.4249 &1.1403 &&0 &0 &&$-$1.4694 &6.4754 &&0 &0 \\
1500&$\phi_1$&$-$0.0045 &0.0010 &&0.0004 &0.0003 &&0.0037 &0.0006 &&0 &0.0002 \\
&$\phi_2$&$-$0.0145 &0.0110 &&0.0017 &0.0016 &&0.0239 &0.0057 &&0.0007 &0.0008 \\
&$\lambda$&0.0755 &0.1913 &&$-$0.0068 &0.0403 &&$-$0.0841 &0.1812 &&$-$0.0019 &0.0463 \\
&$r$&$-$0.1935 &0.5023 &&0 &0 &&$-$0.9999 &4.0159 &&0 &0 \\
\hline
\end{tabular}}
\end{adjustwidth}
\end{table}

Table \ref{r-estimate} shows the D-Ness algorithm and the two estimation method performances of $\hat{r}$ for Models A1 - B4, including the mean,
the percentage of correctly identifying $r$ (${\rm CP}(r)$) across 10000 replications and duration(s), that is, the computing time in seconds.
For the D-Ness algorithm, we set $(\underline{\lambda},\overline{\lambda},L)=(2,6,4)$.
Below, we will compare the three methods in terms of ${\rm CP}(r)$ and duration(s).

In term of the percentage chance of correctly identifying $r$, there is no doubt that the advantage of $\hat{r}_{CML}$ is particularly obvious. In particular, ${\rm CP}(r)$ is above $0.9$ with small sample size ($n=200$) for all models. For the results of $\hat{r}_{CLS}^{(2)}$, although the correct percentage is not high when the sample size is small, it can also reach more than 0.9 (when $R=0$) as the sample size increases ($n=1500$). For $R=1$, simulations show that when the sample size reaches $4000$,
${\rm CP}(r)$ can also reach $0.9$. To save space, the details are not displayed in the table.
At the same time, it is noted that the ${\rm CP}(r)$ of $\hat{r}_{CLS}^{(2)}$ when $R=1$ is obviously not as good as when $R=0$,
which is mainly because the variance difference between the two regimes is not as large as when $R=0$, especially for Model B4, which also explains why the result of threshold estimation for Model B4 is not ideal. The result of the D-Ness algorithm shows that its performance in Models A2 and A4 (B2 and B4) is extremely poor, as we analyzed in Set. 3.3, mainly because $\phi_1=\phi_2$ in these models.

In terms of calculation time (duration(s)), since the D-Ness algorithm finds the threshold value from closed-form expressions, it will be significantly faster than the other two methods, but the premise of using this algorithm is that $T_{Wald-E}$ (\ref{wald-e}) rejects the null Hypothesis $H_0^{(1)}$.
Otherwise, as in the previous analysis, it is impossible to find the true threshold value of the BiNB-MTTINAR(1) model.
For $\hat{r}_{CLS}^{(2)}$, although its calculation speed is not as good as the D-Ness algorithm, it is faster than $\hat{r}_{CML}$, especially for a large sample size.

Based on the above analysis, we make the following recommendations. For a small sample size, our optimal choice for estimating the threshold value is $\hat{r}_{CML}$. For a large sample size, if we can test $\phi_1\neq\phi_2$, the D-Ness algorithm is the optimal choice; otherwise, $\hat{r}_{CLS}^{(2)}$ is the best.
\begin{table}           
\begin{adjustwidth}{0.5cm}{1cm}
\scriptsize
\caption{The performances of $\hat{r}$ for models A1 - B4}
\label{r-estimate}                    
\renewcommand\arraystretch{0.8}        
\vspace{-3mm}                       
{\tabcolsep0.06in                     
\begin{tabular}{*{13}{l}}
\hline
&&\multicolumn{3}{c}{D-Ness}&&\multicolumn{3}{c}{CLS estimator $\hat{r}_{CLS}^{(2)}$}&&\multicolumn{3}{c}{CML estimator $\hat{r}_{CML}$}\\\cline{3-5}\cline{7-9}\cline{11-13}
Model&$n$&Mean&${\rm CP}(r)$&Duration(s)&&Mean&${\rm CP}(r)$&Duration(s)&&Mean&${\rm CP}(r)$&Duration(s)\\\hline
A1&200&4.7590 &0.4443 &2.9870 &&5.1117 &0.5231 &1044.5860 &&4.0133 &0.9826 &3504.3890 \\
&500&4.4779 &0.6627 &3.5810 &&4.5716 &0.7100 &1303.9650 &&4.0001 &0.9999 &7768.6120 \\
&800&4.2766 &0.8043 &4.7590 &&4.3502 &0.8148 &1535.6430 &&4.0000 &1.0000 &13009.4190 \\
&1500&4.0912 &0.9362 &5.9880 &&4.1370 &0.9196 &1359.1520 &&4.0000 &1.0000 &20730.0420 \\
A2&200&5.7958 &0.1440 &3.1470 &&5.6532 &0.4444 &1251.7120 &&4.0399 &0.9470 &4650.2890 \\
&500&5.5725 &0.1948 &4.2800 &&4.8042 &0.6927 &1730.1670 &&4.0012 &0.9980 &10875.8460 \\
&800&5.4216 &0.2337 &4.5590 &&4.4330 &0.8204 &1506.4450 &&4.0001 &0.9999 &16219.0090 \\
&1500&5.2820 &0.2696 &10.8690 &&4.1195 &0.9403 &2886.6380 &&4.0000 &1.0000 &34187.4860 \\
A3&200&13.5680 &0.1185 &10.8900 &&9.2616 &0.4839 &5472.8070 &&7.0038 &0.9236 &29578.2180 \\
&500&12.8048 &0.2679 &9.0360 &&7.7692 &0.7708 &3700.9410 &&6.9990 &0.9964 &61768.0290 \\
&800&12.3470 &0.3544 &13.0000 &&7.3615 &0.8759 &4452.3240 &&6.9998 &0.9998 &103513.5660 \\
&1500&11.4760 &0.4820 &15.7860 &&7.0702 &0.9701 &3738.1580 &&7.0000 &1.0000 &178727.1130 \\
A4&200&15.0163 &0.1129 &5.5600 &&14.0816 &0.4582 &2890.8690 &&12.0320 &0.9489 &24881.2080 \\
&500&14.7580 &0.1382 &7.4260 &&13.0950 &0.6924 &3257.2870 &&12.0008 &0.9980 &63494.5690 \\
&800&14.8073 &0.1367 &9.9120 &&12.6005 &0.8039 &3829.1370 &&12.0002 &0.9998 &105284.7910 \\
&1500&14.6125 &0.1567 &11.5210 &&12.2046 &0.9230 &2939.0570 &&12.0000 &1.0000 &175887.6020 \\
\hline
&&\multicolumn{3}{c}{D-Ness}&&\multicolumn{3}{c}{CLS estimator $\hat{r}_{CLS}^{(2)}$}&&\multicolumn{3}{c}{CML estimator $\hat{r}_{CML}$}\\\cline{3-5}\cline{7-9}\cline{11-13}
Model&$n$&Mean&${\rm CP}(r)$&Duration(s)&&Mean&${\rm CP}(r)$&Duration(s)&&Mean&${\rm CP}(r)$&Duration(s)\\\hline
B1&200&3.7723 &0.5411 &4.8820 &&2.8790 &0.3895 &1723.3820 &&4.0225 &0.9696 &6518.4220 \\
&500&3.8636 &0.7495 &8.1240 &&3.1194 &0.4834 &2661.0870 &&4.0006 &0.9992 &16627.9210 \\
&800&3.9045 &0.8526 &9.6060 &&3.2629 &0.5378 &2685.2930 &&4.0000 &1.0000 &25904.2070 \\
&1500&3.9507 &0.9375 &11.4370 &&3.4517 &0.6242 &2313.6080 &&4.0000 &1.0000 &40997.9390 \\
B2&200&3.7774 &0.2547 &3.1480 &&3.0103 &0.4506 &1016.0670 &&4.0146 &0.9368 &5479.4990 \\
&500&3.9586 &0.2752 &3.9030 &&3.2350 &0.5646 &1099.0420 &&4.0000 &0.9970 &12834.6300 \\
&800&4.0485 &0.2899 &6.0270 &&3.3846 &0.6339 &1502.6660 &&4.0000 &1.0000 &22731.9480 \\
&1500&4.1099 &0.3177 &12.0070 &&3.5882 &0.7365 &2389.2140 &&4.0000 &1.0000 &44905.2730 \\
B3&200&6.6826 &0.7092 &8.3850 &&6.0259 &0.5457 &4009.6040 &&6.9938 &0.9449 &19935.8860 \\
&500&6.9143 &0.9143 &9.2810 &&6.3792 &0.7040 &3540.5920 &&6.9998 &0.9986 &45079.4920 \\
&800&6.9670 &0.9677 &9.4190 &&6.5751 &0.7859 &2842.2780 &&7.0000 &1.0000 &64377.6220 \\
&1500&6.9950 &0.9950 &15.3160 &&6.8065 &0.8981 &3476.0420 &&7.0000 &1.0000 &123651.9970 \\
B4&200&8.9524 &0.1073 &11.2860 &&9.6522 &0.3374 &4611.1180 &&11.9939 &0.9328 &33431.1240 \\
&500&8.9987 &0.1136 &13.2190 &&10.1836 &0.4229 &4454.6210 &&11.9996 &0.9974 &81908.7640 \\
&800&9.0778 &0.1245 &16.3990 &&10.5306 &0.4931 &4685.9790 &&12.0000 &1.0000 &136090.3950 \\
&1500&9.1444 &0.1272 &21.3990 &&11.0001 &0.6095 &4452.9270 &&12.0000 &1.0000 &225375.6450 \\
\hline
\end{tabular}
}
\end{adjustwidth}
\end{table}

\subsection{Size and Power of the Wald test}
In this section, we will evaluate the performance of the test statistics $T_{Wald-E}$ and $T_{Wald-Var}$. For the sake of readability, we introduce the following two INAR processes that are used in simulations:
\begin{flalign}
&\text{\emph{INAR(1)-Poisson:}~~~}X_t=\alpha_{1}\circ X_{t-1}+Z_{1,t},&\label{inar-p}\\
&\text{\emph{INAR(1)-Geo:}~~~}X_t=\alpha_{2}\ast X_{t-1}+Z_{2,t}.&\label{inar-g}
\end{flalign}
where ``$\alpha_{1}\circ$", ``$\alpha_{2}\ast$", $Z_{1,t}$, $Z_{2,t}$ are the same as in Definition \ref{BiNB}.

For each simulation, the empirical sizes and empirical powers, the relative frequency of simulated sample paths leading to a rejection of the null (rejection rate), are calculated at the nominal level of 0.05 (the associated critical values are 3.8415 and 5.991). The repetition number is 10000, and the sample sizes $n=200, 500, 800, 1000$ are used to calculate the empirical sizes and the sample sizes $n=200, 500, 1000, 2000$ are used to calculate the empirical powers.
We begin by calculating empirical sizes using observations produced by models (\ref{inar-p}) and (\ref{inar-g}), with the following parameter configurations taken into account:\\
Model(I-P${_1}$): $(\alpha_{1},\lambda)=(0.2,6)$;~~Model(I-P${_2}$): $(\alpha_{1},\lambda)=(0.4,5)$;~~Model(I-P${_3}$): $(\alpha_{1},\lambda)=(0.5,5)$;\\
Model(I-G${_1}$): $(\alpha_{2},\lambda)=(0.4,5)$;~~Model(I-G${_2}$): $(\alpha_{2},\lambda)=(0.5,4)$;~~Model(I-G${_3}$): $(\alpha_{2},\lambda)=(0.6,5)$;\\
Then, we calculate empirical powers using observations generated through the BiNB-MTTINAR(1) model (\ref{Mymodel}) with the following parameter configurations:\\
Model(B-M${_1}$): $(\phi_1,\phi_2,\lambda,r,R)=(0.4,0.2,6,6,0)$;~~Model(B-M${_4}$): $(\phi_1,\phi_2,\lambda,r,R)=(0.4,0.2,6,6,1)$;\\
Model(B-M${_2}$): $(\phi_1,\phi_2,\lambda,r,R)=(0.4,0.4,6,6,0)$;~~Model(B-M${_5}$): $(\phi_1,\phi_2,\lambda,r,R)=(0.4,0.4,6,6,1)$;\\
Model(B-M${_3}$): $(\phi_1,\phi_2,\lambda,r,R)=(0.3,0.6,5,7,0)$;~~Model(B-M${_6}$): $(\phi_1,\phi_2,\lambda,r,R)=(0.3,0.6,5,7,1)$;\\
Among them, B-M${_2}$ and B-M${_4}$ models are the special cases in the BiNB-MTTINAR(1) model, with $\phi_1=\phi_2$, and can be used to calculate the empirical sizes of $T_{Wald-E}$ and the empirical powers of $T_{Wald-Var}$.
Illustrative results concerning the size and power are shown in Tables \ref{test-size} and \ref{test-pow}.
\begin{table}
\centering
\caption{Empirical Sizes of $T_{Wald-E}$ and $T_{Wald-Var}$ at level 0.05}
\label{test-size}                    
\renewcommand\arraystretch{0.8}        
\vspace{-3mm}                       
{\tabcolsep0.09in                     
\begin{tabular}{lcccclccc}
\hline
Model&$n$&$T_{Wald-E}$&$T_{Wald-Var}$&&Model&$n$&$T_{Wald-E}$&$T_{Wald-Var}$\\\hline
I-P${_1}$&200&0.0293 &0.0350 &&I-G${_1}$&200&0.0273 &0.0425 \\
&500&0.0345 &0.0497 &&&500&0.0525 &0.0450 \\
&800&0.0447 &0.0548 &&&800&0.0519 &0.0494 \\
&1000&0.0485 &0.0522 &&&1000&0.0490 &0.0481 \\\hline
I-P${_2}$&200&0.0525 &0.0048 &&I-G${_2}$&200&0.0523 &0.0356 \\
&500&0.0502 &0.0250 &&&500&0.0513 &0.0490 \\
&800&0.0484 &0.0440 &&&800&0.0505 &0.0492 \\
&1000&0.0483 &0.0514 &&&1000&0.0532 &0.0506 \\\hline
I-P${_3}$&200&0.0543 &0.0197 &&I-G${_3}$&200&0.0504 &0.0337 \\
&500&0.0561 &0.0449 &&&500&0.0521 &0.0527 \\
&800&0.0526 &0.0577 &&&800&0.0506 &0.0549 \\
&1000&0.0503 &0.0521 &&&1000&0.0526 &0.0568 \\\hline
B-M${_2}$&200&0.0287 &&&B-M${_5}$&200&0.0242 &\\
&500&0.0490 &&&&500&0.0562 &\\
&800&0.0501 &&&&800&0.0556 &\\
&1000&0.0504 &&&&1000&0.0542 &\\
\hline
\end{tabular}
}
\end{table}

\begin{table}
\centering
\caption{Empirical Power of $T_{Wald-E}$ and $T_{Wald-Var}$ at level 0.05}
\label{test-pow}                    
\renewcommand\arraystretch{0.8}        
\vspace{-3mm}                       
{\tabcolsep0.09in                     
\begin{tabular}{lcccclccc}
\hline
Model&$n$&$T_{Wald-E}$&$T_{Wald-Var}$&&Model&$n$&$T_{Wald-E}$&$T_{Wald-Var}$\\\hline
B-M${_1}$&200&0.2889&0.3448&&B-M${_4}$&200&0.0585&0.0807\\
&500&0.5284&0.6524&&&500&0.2364&0.2415\\
&1000&0.8171&0.8396&&&1000&0.3988&0.5538\\
&2000&0.9778&0.9512&&&2000&0.6301&0.8963\\\hline
B-M${_2}$&200&&0.3904&&B-M${_5}$&200&&0.0501\\
&500&&0.7228&&&500&&0.1274\\
&1000&&0.9148&&&1000&&0.3216\\
&2000&&0.9836&&&2000&&0.6976\\\hline
B-M${_3}$&200&0.5983&0.2373&&B-M${_6}$&200&0.4881&0.1085\\
&500&0.9204&0.5676&&&500&0.8759&0.3782\\
&1000&0.9971&0.8762&&&1000&0.9948&0.7810\\
&2000&1.0000&0.9856&&&2000&1.0000&0.9821\\
\hline
\end{tabular}
}
\end{table}
As can be seen from Tables \ref{test-size} and \ref{test-pow}, both $T_{Wald-E}$ and $T_{Wald-Var}$ achieve satisfactory performances, and the empirical sizes are closer to the significance level of 0.05 as the sample size increases.
In particular, the empirical sizes of $T_{Wald-E}$ approach the $0.05$ level significantly faster, because the estimation of conditional variance can be regarded as the two-step CLS estimation based on $\hat{\bm{\theta}}_{CLS}$, and its convergence rate must be slower than $\hat{\bm{\theta}}_{CLS}$.
For the Models B-M${_2}$ and B-M${_4}$, we expect that the true conclusion can be obtained through our tests: the existence of the segmented structure. However, as can be seen from Tables \ref{test-size} and \ref{test-pow}, $T_{Wald-E}$ does not meet this expectation,
which is consistent with our argument in Sect. 3.4.2 that $T_{Wald-E}$ cannot be used to support the conclusion that the segmented structure does not exist.
Based on the above analysis, we can draw the following conclusions. In practice, we can use $T_{Wald-E}$ to test first, and if $T_{Wald-E}$ rejects the null hypothesis, the piecewise structure exists. If $T_{Wald-E}$ supports the null hypothesis, we need to check again using $T_{Wald-Var}$ to detect the existence of a piecewise structure.
\section{Real Data Example}
In this section, we discuss possible applications of the introduced BiNB-MTTINAR(1) model.
We consider the datasets from The Forecasting Principles site (\url{http://www.forecastingprinciples.com}), in the section about Crime - Pittsburgh police car beat data. The data set recorded the number of monthly occurrences of 36 types of crime in 66 police departments in Pittsburgh, and has been partially analyzed by \cite{Zhu2012}, \cite{Zhangetal2020} and \cite{Zhangetal2022}, among others.
For this example, we focus on the counts of criminal mischief in the $14$th police car beat in Pittsburgh. The data consist of 144 observations, starting in January 1990 and ending in December 2001.
\begin{figure}[H]
\begin{adjustwidth}{-1.8cm}{1cm}
    \includegraphics[width=7.5in,height=2.6in]{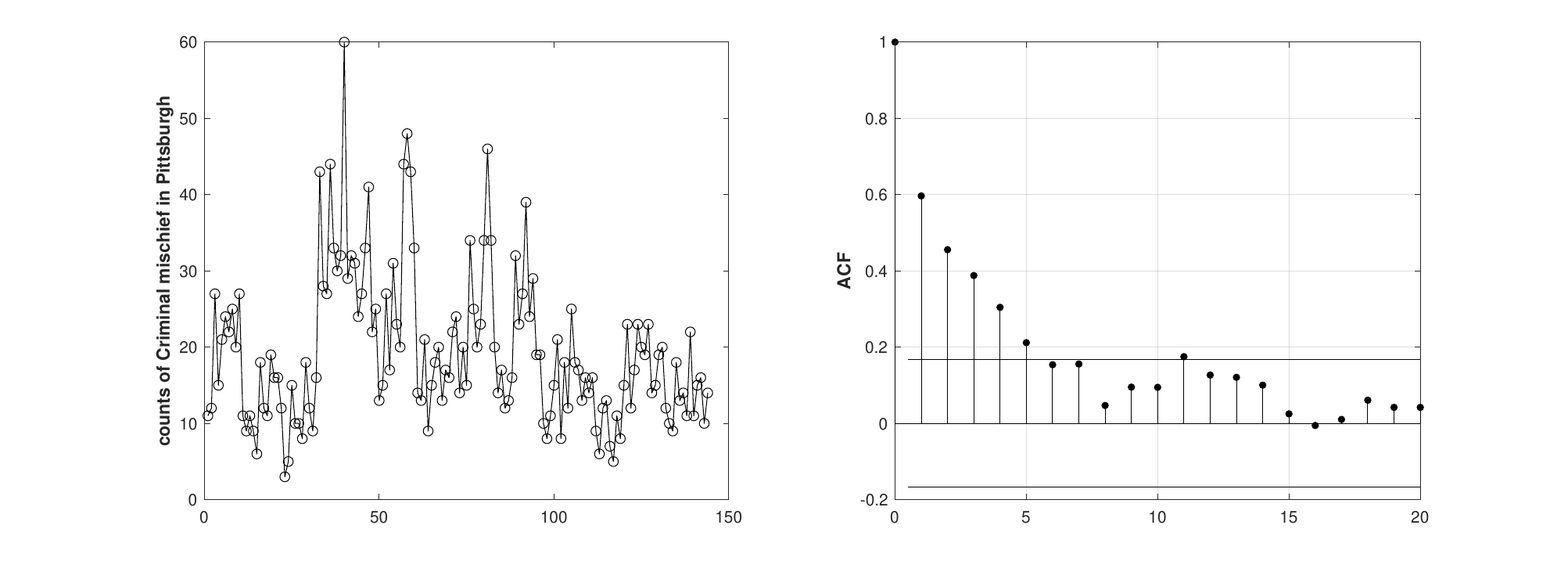}
\end{adjustwidth}
\vspace{-8mm}
\caption{Sample path and ACF of the monthly counts of Criminal mischief in Pittsburgh \protect\\(from January 1990 and ending in December 2001).}
\label{sample_CMIS}
\end{figure}
Figure \ref{sample_CMIS} shows the sample path and the sample autocorrelation (ACF) of the observations.
We start with fitting the BiNB-MTINAR(1) (when $R=0$) model to the data (later, we also consider the more general SETINAR models). We estimate the model parameters for threshold values of $r\in\{9,..., 33\}$, where $9$ and $33$ are the $10$th and $90$th quantiles of the data.
After minimizing (\ref{minr_cml}), we decide to consider a model with a threshold value $\hat{r}_{CML}=11$. Then, we applied the Wald test (\ref{wald-e}) to check whether such a nonlinear model is appropriate for the data. We obtain $T_{Wald-E}=3.9636$, while our critical value on a $0.05$ level is 3.8415, so we have to reject the null hypothesis $\phi_1=\phi_2$.
It should be noted that if the $T_{Wald-E}$ obtained is less than 3.8415, we need to further apply $T_{Wald-Var}$ (\ref{wald-var}) to check for the existence of a piecewise structure.

Next, we use the BiNB-MTINAR(1) model and the following integer-valued threshold autoregressive models to fit the criminal mischief incident dataset and compare different models via the AIC and BIC.
\begin{itemize}
  \setlength{\itemsep}{1pt}
  \setlength{\parskip}{0pt}
  \setlength{\parsep}{0pt}
  \item SETINAR(2,1)~model~(\cite{Monteiro2012}).
  \item NBTINAR(1) model (\cite{Yang2018}).
  \item RCTINAR(1) model (\cite{LiY2018}).
\end{itemize}
For each of the above models, the CML of the parameters and the threshold $r$,
the standard error (SE) of $\hat{\bm{\theta}}_{CML}$, the root mean square of differences between observations and forecasts (RMS), and the AIC and BIC values are given. Among them, the standard error for the CML estimator can be obtained as the square roots of the elements in the diagonal of the inverse of the negative Hessian of the log-likelihood calculated at the CML estimates. RMS is defined as follows
$${\rm RMS}=\sqrt{\frac{1}{n-1}\sum\limits_{t=2}^{n}(X_t-\hat{\phi}_{1}X_{t-1}I_{1,t}^R-\hat{\phi}_{2}X_{t-1}I_{2,t}^R-\hat{\lambda})^2}$$
The fitting results are summarized in Table \ref{aicbic}.
As seen from the results presented in Table \ref{aicbic}, the parameter $\alpha_1$ estimators for the SETINAR(2,1) and NBTINAR(1) models are poor since they are close to the boundaries 0 and 1 and are not significant regarding the approximated standard errors. Therefore, these two models are not suitable for fitting this dataset.
Furthermore, when comparing the SETINAR(2,1) and RCTINAR(1) models, we find that all of their parameter estimations are similar. The outcome of RCTINAR(1) is very superior to that of SETINAR when taking AIC and BIC into account as information criteria.
This conclusion is consistent with the finding in \cite{LiY2018}, which states that SETINAR(2,1) is a special case of the RCTINAR(1) model.

\begin{figure}
\begin{adjustwidth}{0.2cm}{1cm}
    \includegraphics[width=6.5in,height=3.8in]{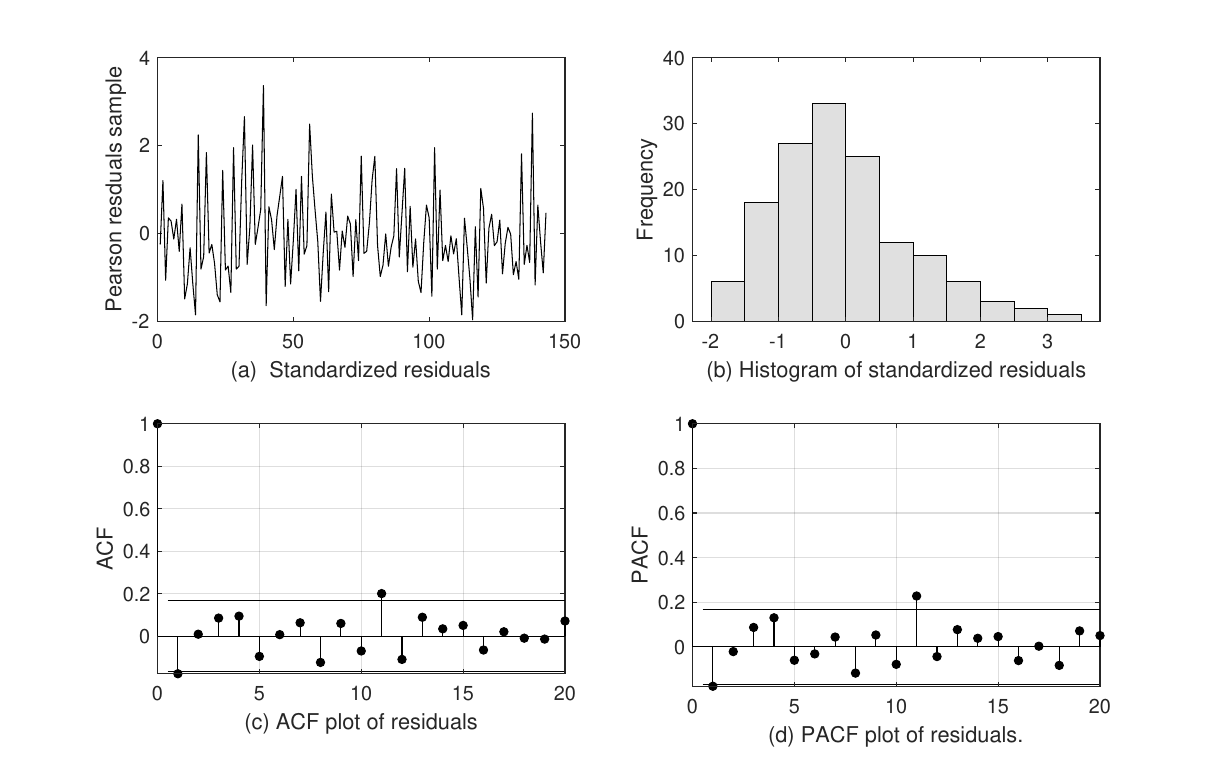}
\end{adjustwidth}
\vspace{-6mm}
\caption{Diagnostic checking plots for the Criminal mischief in Pittsburgh data.}
\label{Pearson_resduals}
\end{figure}

Then, we compute the (standardized) Pearson residuals, Pr$_t(\hat{\bm{\theta}}$) \citep{Weiss2019}, to check if the fitted model is adequate for the data.
\begin{align}
{\rm Pr}_t(\hat{\bm{\theta}})= \frac{X_t-\hat{\phi}_{1}X_{t-1}I_{1,t}^R-\hat{\phi}_{2}X_{t-1}I_{2,t}^R-\hat{\lambda}}{\sqrt{(\hat{\phi}_{1}(1-\hat{\phi}_{1})X_{t-1}+\hat{\lambda})I_{1,t}^R +(\hat{\phi}_{2}(1+\hat{\phi}_2)X_{t-1}+\hat{\lambda}(1+\hat{\lambda}))I_{2,t}^R}}.
\end{align}
After computing, the mean and the variance of Pr$_t(\hat{\bm{\theta}}$) are $-0.0483$ and $1.0653$, which are close to 0 and 1, indicating that the fitted model is adequate.
Furthermore, Figure \ref{Pearson_resduals} shows the diagnostic checking  plots for our out fitted model,
including (a) Standardized residuals, (b) Histogram of standardized residuals, (c) ACF plot of residuals, and (d) PACF plot of residuals.
As can be roughly seen from the residuals sample plot (Figure \ref{Pearson_resduals}(a)), the estimated residuals indicate that the series is stationary,
and the Pearson residuals samples ACF and PACF have values close to zero, which reveals that our fitted model is suitable.

Finally, all estimators of the BiNB-MTINAR(1) ($R=0$) model are significant regarding the approximated standard errors, and the BiNB-MTINAR(1) ($R=0$) model has the lowest RMS, AIC, and BIC of all models and is the model of choice considering these information criteria.
\begin{table}
\renewcommand\arraystretch{0.89}      
\caption{Fitting results of different models: CML, $\hat{r}_{CML}$, SE, RMS, AIC and BIC}
\label{aicbic}
\vspace{-3mm}                        
\centering                            
{\tabcolsep0.1in                     
\begin{tabular}{*{8}{c}}
  \hline
Model&Para.&CML&SE&$\hat{r}_{CML}$&RMS&AIC&BIC\\\hline
SETINAR(2,1)&$\alpha_1$&0.0000 &3.5494 &14&8.0297 &1140.5533 &1149.4627 \\
&$\alpha_2$&0.3454 &0.0048 &&&&\\
&$\lambda$&14.0482 &0.0010& &&&\\
NBTINAR(1)&$\alpha_1$&1.0000 &0.6167&17 &7.9392 &984.8061 &993.7155 \\
&$\alpha_2$&0.7811 &0.0061& &&&\\
&$v$&3.0000 &0.0014& &&&\\
RCTINAR(1)&$\phi_1$&0.0744&0.0054 &14&7.9888 &1009.0917 &1018.0011 \\
&$\phi_2$&0.3751 &0.0029 &&&&\\
&$\lambda$&13.2755 &0.0004 &&&&\\
BiNB-MTTINAR(1)&$\phi_1$&0.3872&0.0029 &11&7.9261 &974.2542 &983.1636 \\
(R=0)&$\phi_2$&0.6098 &0.0029&&&&\\
&$\lambda$&8.5907 &0.0005& &&&\\
BiNB-MTTINAR(1)&$\phi_1$&0.5508&0.0006 &33 &8.0711 &998.5767 &1007.4862 \\
(R=1)&$\phi_2$&0.6581 &0.0010& &&&\\
&$\lambda$&7.6882 &0.0012& &&&\\
\hline
\end{tabular}
}
\end{table}

A useful and significant problem in time series analysis is the prediction problem.
Under the standard framework, conditional expectation is the frequently used approach for constructing predictors since it can yield predictors with the optimal property in terms of the mean squared error.
However, the conditional expectation method violates data coherence since it can hardly produce integer-valued predictors.
Therefore, we want an appropriate method that can generate predictors with integer values.
\cite{Freeland2004} proposed a workable method that uses the $h$-step-ahead conditional distribution to forecast the future value.
One can obtain the point predictor from the median or the mode of the  $h$-step-ahead conditional distribution.
This method has been employed by some researchers, see \cite{Maiti2017}, \cite{LiY2018}, \cite{Kang2021}.
Learning from this idea, it is possible to calculate the $h$-step-ahead conditional distribution for the BiNB-MTINAR(1) model by taking powers of the transition matrix since we are interested in a Markov chain.
To be more precise, one may calculate the $h$-step-ahead conditional distribution of $X_{t+h}$ given $X_t$ by
\begin{align*}
\mathrm{P}(X_{t+h}=j|X_{t}=i)=[\textbf{P}^{h}]_{i+1,j+1},
\end{align*}
where $\textbf{P}$ denotes the transition matrix defined by the following equation,
\begin{align}\label{2.4}
\mathrm{\mathbf{P}}=(\mathrm{P}_{ij})_{i,j=0,1,2...,}
\end{align}
and $\mathrm{P}_{ij}=\mathrm{P}(X_t=j|X_{t-1}=i)$ is the transition probability, which is given in Equation (\ref{transprob}).
For illustration purposes, we compute the mean and mode of the $h$-step-ahead conditional distribution for the BiNB-MTINAR(1) ($R=0$) model and their corresponding bias and the mean absolute deviation error (MADE) are considered in Table \ref{prect}, where MADE is defined as follows:
\begin{align*}
\text{MADE}=\frac{1}{n-1}\sum\limits_{t=1}^n|\hat{X}_t-X_t|.
\end{align*}
From the result in Table \ref{prect}, the bias and MADE of the proposed predictors are acceptable.
Therefore, based on the above analysis, we can conclude that the BiNB-MTINAR(1) (when $R=0$) model is appropriate for this data set.
\begin{table}[H]
\centering
	\caption{Bias and MADE for $h$-step ahead predictions of the crime data: Criminal mischief }\label{prect}
\begin{tabular}{lrlcrl}
			\hline
			\multirow{2}{*}{$h$}&\multicolumn{2}{c}{Expectation}&&\multicolumn{2}{c}{Mode}\\\cline{2-3}\cline{5-6}
			~&\multicolumn{1}{c}{Bias}&MADE&&\multicolumn{1}{c}{Bias}&MADE\\\hline
1&$-$1.5378&1.5378&&$-$2.0000&2.0000\\
5&1.8176&2.4252&&1.6000&2.4000\\
10&$-$0.6992&2.7403&&$-$1.0000&2.8000\\
20&$-$0.9485&3.1871&&$-$1.0500&3.2500\\
30&$-$0.6581&3.9260&&$-$0.7667&3.8333\\
\hline
\end{tabular}
\end{table}
\section{Conclusions}
This article introduced a new first-order mixture integer-valued threshold autoregressive process based on the binomial and negative binomial thinning operators.
The process is proven to be stationary and ergodic.
We investigated the CLS and the CML techniques for parameter estimation, and the estimators' asymptotic properties are demonstrated.
Two methods are suggested for estimating the unknown threshold parameter $r$, based on the CLS and CML score functions.
Additionally, to check for the existence of piecewise structure, we constructed two types Wald test statistics for conditional expectation and conditional variance parameters, respectively.
We successfully applied the BiNB-MTINAR(1) model to crime time series dataset.
Potential issues for future research include extending the results to multivariate cases and considering mixture threshold INAR models based on generalized thinning operator. These remain topics for future study.\\
~\\
\vspace{3mm} {\Large\bf  Acknowledgements} \vspace{1mm}

The authors thank the Joint Editor, Associate Editor and two reviewers for helpful comments,
which led to a much improved version of the paper.
This work is supported by National Natural Science Foundation of China (No.12171463, 11871028, 11731015, 12001229, 11901053, 12001229).

\vskip 0.2cm
{\bf \Large \hspace{-1.25em}Appendix }

\vskip 0.2cm
\renewcommand\theequation{\arabic{equation}}
{\bf \hspace{-1.6em}Proof of Proposition 2.1}
Considering that the case $R=0$ is similar to $R=1$, we only prove the case $R=0$. It is easy to see that $\{X_t\}_{t \in \mathbb{Z}}$ is a Markov chain with state space $\mathbb{N}_0$. From the expression of transition probabilities defined in (\ref{transprob}), it follows that the chain is  irreducible and aperiodic. Furthermore, to show that $\{X_t\}_{t \in \mathbb{Z}}$ is positive recurrent it is sufficient to prove that~$\sum_{t=1}^{\infty}P^t(0,0)=+\infty$
(since $\{X_t\}_{t \in \mathbb{Z}}$ is irreducible)
with $P^t(x,y):=P(X_t=y|X_0=x)$. For convenience we denote
\begin{equation}
    S_t=\begin{cases}
    0, & \text{if $X_{t-1} {\leq} r$},\nonumber\\
    1, & \text{if $X_{t-1} {>} r$}.\nonumber
    \end{cases}
    ~~~~
    \otimes_t:=\begin{cases}
    \circ, & \text{if $X_{t-1} {\leq} r$},\nonumber\\
    \ast, & \text{if $X_{t-1} {>} r$}.\nonumber
    \end{cases}
    ~~~~U_i=\begin{cases}
    e^{-\lambda}, & i=1,\nonumber\\
    \frac{1}{1+\lambda}, & i=2.\nonumber
    \end{cases}
\end{equation}
Then (\ref{Mymodel}) can be rewritten as
\begin{equation}\label{Mymode2}
    X_{t}=\phi_{S_t+1}\otimes_t X_{t-1}+Z_{S_t+1,t}.
\end{equation}
By iterating (\ref{Mymode2}) $t-1$ times, we have
$$
X_t=\phi_{S_t+1}\otimes_t\phi_{S_{t-1}+1}\otimes_{t-1} \cdots \otimes_{2}\phi_{S_1+1}\otimes_1 X_0
+(\sum_{i=1}^{t-1}\phi_{S_t+1} \otimes_{t} \phi_{S_{t-1}+1} \otimes_{t-1}\cdots \otimes_{i+2} \phi_{S_{i+1}+1}\otimes_{i+1} Z_{S_i+1,i})+Z_{S_t+1,t}.
$$
This allows us to write
\begin{align}
P^t(0,0)
&=P((\sum_{i=1}^{t-1}\phi_{S_t+1} \otimes_{t} \phi_{S_{t-1}+1} \otimes_{t-1}\cdots \otimes_{i+2} \phi_{S_{i+1}+1}\otimes_{i+1} Z_{S_i+1,i})+Z_{S_t+1,t}=0|X_0=0)\nonumber\\
&=P(Z_{S_t+1,t}=0,\phi_{S_t+1}\otimes_t Z_{S_{t-1}+1,t-1}=0,\cdots,
\phi_{S_t+1}\otimes_t\cdots\otimes_3\phi_{S_2+1}\otimes_2 Z_{S_1,1}=0|X_0=0)\nonumber\\
&=\sum_{i_1=1}^2\sum_{i_2=1}^2\cdots\sum_{i_t=1}^2 P(S_1+1=i_1,S_2+1=i_2,\cdots,S_t+1=i_t|X_0=0)\nonumber\\
&~~~~\times P(Z_{S_t+1,t}=0,\phi_{i_t}\otimes_t Z_{i_{t-1}+1,t-1}=0,\cdots,
\phi_{i_t}\otimes_t\cdots\otimes_3\phi_{i_2}\otimes_2 Z_{i_1+1,1}=0|X_0=0)\nonumber
\end{align}
By the definition of the binomial thinning operator ``$\circ$" and the negative binomial thinning operator ``$\ast$",
we have
\begin{equation}\label{p00}
P(\phi_{i}\otimes X=0)=
\begin{cases}
    P(\phi_1\circ X=0)=\sum\limits_{m=0}^{+\infty}(1-\phi_1)^mP(X=m), \text{if $i=1$}, \\
    P(\phi_2\ast X=0)=\sum\limits_{m=0}^{+\infty}(1+\phi_2)^{-m}P(X=m),\text{if $i=2$}.
    \end{cases}
\end{equation}
Obviously, if $X$ is not always equal to $0$, $P(\phi_{i}\otimes X=0)>P(X=0), i=1,2$.
Let $k$ denote the number of the sequence $\{X_t\leq r\}_{t \in \mathbb{Z}}$. By (\ref{p00}) we have
\begin{align}
P^t(0,0)&>\sum_{i_1=1}^2\sum_{i_2=1}^2\cdots\sum_{i_t=1}^2 P(S_1+1=i_1,S_2+1=i_2,\cdots,S_t+1=i_t|X_0=0)\nonumber\\
&~~~~\times U_{i_t}U_{i_{t-1}}\cdots U_{i_{1}}\nonumber\\ 
&=\sum_{i_1=1}^2\sum_{i_2=1}^2\cdots\sum_{i_t=1}^2 P(S_1+1=i_1,S_2+1=i_2,\cdots,S_t+1=i_t|X_0=0)\nonumber\\
&~~~~\times e^{-k\lambda}(\frac{1}{1+\lambda})^{t-k}\nonumber
\end{align}
Since $\lim_{t\rightarrow\infty}e^{-k\lambda}(\frac{1}{1+\lambda})^{t-k}=0$, which implies that $\lim_{t\rightarrow\infty}P^t(0,0)\neq0$.
Therefore, we conclude that $\sum_{t=1}^{\infty}P^t(0,0)=+\infty.$
This proves that $\{X_t\}$ is a positive recurrent Markov chain (and hence ergodic) which ensures the existence of a strictly stationary distribution of BiNB-MTTINAR(1) process. \hspace{4.25in}$\square$

{\bf \hspace{-1.6em}Proof of Proposition 2.2} To show this, we denote $\phi_{\max}=\max\{\phi_1,\phi_2\}$. Compute to see that, under the stationary distribution,
\begin{align}\label{EXt}
{\rm E}(X_{t})
&={\rm E}[(\phi_{1}\circ X_{t-1})I_{1,t}^R]+{\rm E}[(\phi_{2}\ast X_{t-1})I_{2,t}^R]+\lambda\nonumber\\
&={\rm E}[{\rm E}((\phi_{1}\circ X_{t-1})I_{1,t}^R|X_{t-1})]+{\rm E}[{\rm E}((\phi_{2}\ast X_{t-1})I_{2,t}^R|X_{t-1})]+\lambda\nonumber\\
&\leq\phi_{\max}{\rm E}(X_{t-1})+\lambda\nonumber\\
&\leq\cdots\nonumber\\
&\leq(\phi_{\max})^{t}{\rm E}(X_{0})+\lambda\sum_{i=0}^{t-1}(\phi_{\max})^i\leq\infty,
\end{align}
where $\phi_{\max}=\max\{\phi_1, \phi_2\}$.\\
Similarly, we have
\begin{align*}
{\rm E}(X_t^2)&=\left[{\rm E}(\phi_{1}\circ X_{t-1})^2+2{\rm E}(\phi_{1}\circ X_{t-1}){\rm E}(Z_{1,t})+{\rm E}(Z_{1,t})^2\right]I_{1,t}^R\\
&+\left[{\rm E}(\phi_{2}\ast X_{t-1})^2+2{\rm E}(\phi_{2}\ast X_{t-1}){\rm E}(Z_{2,t})+{\rm E}(Z_{2,t})^2\right]I_{2,t}^R\\
&=\left[(\phi_1-\phi_1^2+2\phi_1\lambda){\rm E}(X_{t-1})+\phi_1^2{\rm E}(X_{t-1}^2)+\lambda+\lambda^2\right]I_{1,t}^R\\
&+\left[(\phi_2+\phi_2^2+2\phi_2\lambda){\rm E}(X_{t-1})+\phi_2^2{\rm E}(X_{t-1}^2)+\lambda+2\lambda^2\right]I_{2,t}^R\\
&\leq u_{\max}{\rm E}(X_{t-1})+\phi_{\max}^2{\rm E}(X_{t-1}^2)+\lambda+2\lambda^2
\end{align*}
where $u_{\max}=\max\{(\phi_1-\phi_1^2+2\phi_1\lambda),(\phi_2+\phi_2^2+2\phi_2\lambda)\}$.\\
If $t=1$, ${\rm E}(X_1^2)\leq u_{\max}{\rm E}(X_0)+\phi_{\max}^2{\rm E}(X_0^2)+\lambda+2\lambda^2<\infty$.\\
Else if $t \geq 2$,
\begin{align}\label{EXt2}
{\rm E}(X_t^2)&\leq \sum_{i=0}^{t-1}u_{\max}\phi_{\max}^{t-1+i}{\rm E}(X_{0})+\lambda\sum_{i=0}^{t-2}u_{\max}\phi_{\max}^{t-2+i}+(\lambda+2\lambda^2)\sum_{i=0}^{t-1}\phi_{\max}^{2i}\\
&<\infty.\nonumber
\end{align}
Some similar but tedious calculation shows that $E(X_t^3) \leq \infty$. Combining (\ref{EXt}) and (\ref{EXt2}),
one can see that ${\rm E}(X_t^k) < \infty$ for $k = 1, 2, 3$.\hspace{2.96in}$\square$

{\bf \hspace{-1.6em}Proof of Proposition 2.3} The results (\rmnum{1}) to (\rmnum{3}) are straightforward to verify. We give the proof of (\rmnum{4}) only.\\
(\rmnum{4}) The variance of $X_t$ is given by
\begin{align}\label{varX}
{\rm Var}(X_{t})
&={\rm Var}[I_{1,t}^R(\phi_{1}\circ X_{t-1}+Z_{1,t})]+{\rm Var}[I_{2,t}^R(\phi_{2}\ast X_{t-1}+Z_{2,t})]\nonumber\\
&+2{\rm Cov}\big(I_{1,t}^R(\phi_{1}\circ X_{t-1}+Z_{1,t}),I_{2,t}^R(\phi_{2}\ast X_{t-1}+Z_{2,t})\big)\nonumber\\
&=\Rmnum{1}+\Rmnum{2}+\Rmnum{3}.
\end{align}
A direct calculation shows
\begin{align}\label{varpart1}
\Rmnum{1}&={\rm Var}\left[{\rm E}(I_{1,t}^R(\phi_{1}\circ X_{t-1}+Z_{1,t})|X_{t-1})\right]+
{\rm E}\left[{\rm Var}(I_{1,t}^R(\phi_{1}\circ X_{t-1}+Z_{1,t})|X_{t-1})\right]\nonumber\\
&=\phi_1^2{\rm Var}(I_{1,t}^RX_{t-1})+q\lambda+{\rm E}(I_{1,t}^R(\phi_{1}(1-\phi_1)X_{t-1}))\nonumber\\
&=\phi_1^2{\rm Var}(I_{1,t}^RX_{t-1})+q\lambda+q\phi_1(1-\phi_1)\mu_1\nonumber\\
&=\phi_1^2{\rm E}(I_{1,t}^RX_{t-1}^2)-\phi_1^2q^2\mu_1^2+q\lambda+q\phi_1(1-\phi_1)\mu_1\nonumber\\
&=q(\phi_1^2\sigma_1^2+\phi_1(1-\phi_1)\mu_1)+q(1-q)\phi_1^2\mu_1^2+q\lambda
\end{align}
Similarly, we have
\begin{align}\label{varpart2}
\Rmnum{2}&={\rm Var}\left[{\rm E}(I_{2,t}^R(\phi_{2}\ast X_{t-1}+Z_{2,t})|X_{t-1})\right]+
{\rm E}\left[{\rm Var}(I_{2,t}^R(\phi_{2}\ast X_{t-1}+Z_{2,t})|X_{t-1})\right]\nonumber\\
&=\phi_2^2{\rm Var}(I_{2,t}^RX_{t-1})+(1-q)\lambda(1+\lambda)+{\rm E}(I_{2,t}^R(\phi_{2}(1+\phi_2)X_{t-1}))\nonumber\\
&=\phi_2^2{\rm E}(I_{2,t}^RX_{t-1}^2)-\phi_2^2(1-q)^2\mu_2^2+(1-q)\lambda(1+\lambda)+(1-q)\phi_2(1+\phi_2)\mu_2\nonumber\\
&=(1-q)(\phi_2^2\sigma_2^2+\phi_2(1+\phi_2)\mu_2)+q(1-q)\phi_2^2\mu_2^2+(1-q)\lambda(1+\lambda)
\end{align}
and
\begin{align}\label{varpart3}
\Rmnum{3}&=2{\rm Cov}(I_{1,t}^R(\phi_{1}\circ X_{t-1}+Z_{1,t}),I_{2,t}^R(\phi_{2}\ast X_{t-1}+Z_{2,t}))\nonumber\\
&=-2q(1-q)(\phi_1\mu_1+\lambda)(\phi_2\mu_2+\lambda).
\end{align}
Then, (\rmnum{4}) follows by replacing (\ref{varpart1}), (\ref{varpart2}) and (\ref{varpart3}) in (\ref{varX}) and some algebra.\\
(\rmnum{5}):
For the autocovariance ${\rm Cov}(X_t,X_{t+h})$, when $h=1$,
\begin{align*}
{\rm Cov}(X_t,X_{t+1})&={\rm Cov}[X_t,{\rm E}(X_{t+1}|X_{t})]\\
&={\rm Cov}\{X_t,[\phi_{1}X_t+\lambda]I_{1,t+1}^R+[\phi_{2}X_t+\lambda]I_{2,t+1}^R\}\\
&=\sum_{s=1}^2[\phi_s{\rm Cov}(X_t,I_{s,t+1}^RX_t)],
\end{align*}
where
\begin{align*}
{\rm Cov}(X_t,I_{s,t+1}^RX_t)&={\rm E}(I_{s,t+1}^RX_t X_t)-{\rm E}(I_{s,t+1}^RX_t){\rm E}(X_t)=p_s(\sigma_s^2+\mu_s^2)-p_s\mu_s{\rm E}(X_t),
\end{align*}
Then,
\begin{align*}
{\rm Cov}(X_t,X_{t+1})=\sum_{s=1}^2\left\{\phi_sp_s[(\sigma_s^2+\mu_s^2)-\mu_s{\rm E}(X_t)]\right\}=\sum_{s=1}^2\phi_sp_s\gamma_{0}^{(s)},
\end{align*}
when $h>1$, there is
\begin{align*}
{\rm Cov}(X_t,X_{t+h})&={\rm Cov}[X_t,{\rm E}(X_{t+h}|X_{t+h-1})]\\
&={\rm Cov}\{X_t,(\phi_{1}X_{t+h-1}+\lambda)I_{1,t+h}^R+(\phi_{2}X_{t+h-1}+\lambda)I_{2,t+h}^R\}\\
&=\sum_{s=1}^2\phi_{s}[{\rm Cov}(X_t,I_{s,t+h}^RX_{t+h-1})],
\end{align*}
where
\begin{align*}
{\rm Cov}(X_t,I_{s,t+h}^RX_{t+h-1})&=p_s{\rm Cov}(X_t,X_{t+h-1}|X_{t+h-1}\leq r)=p_s\gamma_{h-1}^{(s)},
\end{align*}
Then, we obtain
\begin{align*}
{\rm Cov}(X_t,X_{t+h})=\sum_{s=1}^2\phi_sp_s\gamma_{h-1}^{(s)}.
\end{align*}
Thus, the autocorrelation function $\rho(h)=[\sum_{s=1}^2\phi_sp_s\gamma_{h-1}^{(s)}]\setminus{\rm Var}(X_t)$.
\hspace{1.36in}$\square$

{\bf \hspace{-1.6em}Proof of Theorems \ref{cmlzt}}
We first prove the consistency of $\hat{\bm{\theta}}_{CML}$. Let $\bm{\Theta}$ be the parametric space: $(0,1)^2\times(0,+\infty)$. From (\ref{transprob}), $\ell (\bm{\theta})$ is a measurable function of $X_t$ for all  $\bm{\theta}\in\bm{\Theta}$, and $\partial \ell (\bm{\theta})/\partial \bm{\theta}$ exists and is continuous in an open neighborhood $N_1(\bm{\theta}_0)$ of $\bm{\theta}_0$. Therefore, in order to prove the existence and consistency of the
CML estimators, it is sufficient to show that assumption (c) of Theorems 4.1.2 in \cite{Ame1985} holds.
Note that $\ell_t \bm{(\theta)}$ is continuous in an open and convex neighborhood $N_2(\bm{\theta}_0)$, thus there at least exists a point $\bm{\theta}_1\in N_2(\bm{\theta}_0)$ such that $l_t(\bm{\theta})$ attains the maximum value at $\bm{\theta}_1$, i.e.,
\begin{align}\label{consis1}
\mathrm{E}\left(\sup_{\bm{\theta}\in N_2(\bm{\theta}_0)}l_t(\bm{\theta})\right)=\mathrm{E}\left[\log P(X_t|\mathcal{F}_{t-1})\right]_{\bm{\theta}_1}\leq\log [\mathrm{E}\left(P(X_t|\mathcal{F}_{t-1})\right)_{\bm{\theta}_1}]<\infty.
\end{align}
From Proposition \ref{stationarity}, $\{X_t\}$ is stationary and ergodic, it follows $\frac{1}{n}\sum_{t=1}^n l_t(\bm{\theta})\rightarrow\mathrm{E}l_t(\bm{\theta})$ in probability as $n\rightarrow\infty$. By Jensen's inequality, we have
\begin{align}\label{consis2}
\mathrm{E}(l_t(\bm{\theta})-l_t(\bm{\theta}_0))=\mathrm{E}\left[\log\frac{P(X_t|\mathcal{F}_{t-1})_{\bm{\theta}}}{P(X_t|\mathcal{F}_{t-1})_{\bm{\theta}_0}}\right]
\leq\log\left[\mathrm{E}\frac{P(X_t|\mathcal{F}_{t-1})_{\bm{\theta}}}{P(X_t|\mathcal{F}_{t-1})_{\bm{\theta}_0}}\right]=0.
\end{align}
From (\ref{consis1}) and (\ref{consis2}), $\mathrm{E}(\ell_t \bm{(\theta)})$ is a strict local maximum at $\bm{\theta}_0$.
Hence, the assumption (c) is true, the proof of consistency is complete.

To prove asymptotically, we perform Taylor expansion of the score vector around $\bm{\theta}_0$.
\begin{align}\label{taylor}
\bm{0}=\frac{1}{\sqrt{n}}\frac{\partial{l(\hat{\bm{\theta}}_{CML})}}{\partial\bm{\theta}}=\frac{1}{\sqrt{n}}\frac{\partial{l(\bm{\theta}_0)}}{\partial\bm{\theta}}
+\left(\frac{1}{n}\frac{{\partial^2{l(\bm{\theta}^*)}}}{\partial{\bm{\theta}}\partial{\bm{\theta}^{\mathrm{T}}}}\right)\sqrt{n}(\hat{\bm{\theta}}_{CML}-\bm{\theta}_0),
\end{align}
where $\bm{\theta}^*$ lies in between $\hat{\bm{\theta}}_{CML}$
and $\bm{\theta}_0$. According to Theorem 4.1.3 in \cite{Ame1985}, we divide the proof into the following four steps.\\
$\textbf{Step 1:}$ It is easy to see $\mathrm{E}\dfrac{\partial{l_t(\bm{\theta}_0)}}{\partial\bm{\theta}}=\bm{0}$, thus $Cov\left(\dfrac{\partial{l_t(\bm{\theta}_0)}}{\partial\bm{\theta}}\right)=\mathrm{E}\left[\left(\dfrac{\partial{l_t(\bm{\theta}_0)}}{\partial\bm{\theta}}\right)\left(\dfrac{\partial{l_t(\bm{\theta}_0)}}{\partial\bm{\theta}}\right)^{\mathrm{T}}\right]$. Using the ergodic theorem,
$$\frac{1}{n}\frac{\partial{l(\bm{\theta}_0)}}{\partial\bm{\theta}}\rightarrow \mathrm{E}\left(\frac{1}{P(X_t|\mathcal{F}_{t-1})}\frac{\partial P(X_t|\mathcal{F}_{t-1})}{\partial\bm{\theta}}\right)_{\bm{\theta}_0}~~~\text{in probability one.}$$
Using the martingale central limit theorem and the Cram$\acute{e}$r-Wold device, We can get that
\begin{align}
\dfrac{1}{\sqrt{n}}\dfrac{\partial{l(\bm{\theta}_0)}}{\partial\bm{\theta}}\overset{d}{\longrightarrow}N(\bm{0},\bm{I}(\bm{\theta_0})), \end{align}
where $\bm{I}(\bm{\theta_0})=\mathrm{E}\left[\frac{\partial l_t(\bm{\theta})}{\partial \bm{\theta}}\frac{\partial l_t(\bm{\theta})}{\partial \bm{\theta}^{\mathrm{T}}}\right]_{\bm{\theta_0}}$.\\
$\textbf{Step 2:}$ From (\ref{zygl}), all the partial derivatives $\frac{\partial l(\bm{\theta})}{\partial {\theta}_i}$ exist and three times continuous differentiable in $\bm{\Theta}$, thus $\frac{\partial^2{l(\bm{\theta})}}{\partial {\theta}_i\partial {\theta}_j}$ exists and is continuous in an open, convex neighborhood of $\bm{\theta}_0$ ($\bm{\theta}_2\in N(\bm{\theta}_0)$).\\
$\textbf{Step 3:}$ From Step 2, there at least exists a point $\bm{\theta}_2\in N(\bm{\theta}_0)$ such that $\frac{\partial^2{l(\bm{\theta})}}{\partial {\theta}_i\partial {\theta}_j}$ attains the maximum value at $\bm{\theta}_2$, i.e.,
$$\mathrm{E}\left[\sup_{\bm{\theta}\in N(\bm{\theta}^0)}\left(\frac{\partial^2{l(\bm{\theta})}}{\partial {\theta}_i\partial {\theta}_j}\right)\right]=\mathrm{E}\left(\frac{\partial^2{l(\bm{\theta})}}{\partial {\theta}_i\partial {\theta}_j}\right)_{\bm{\theta}_2}<\infty.$$
For convenience, we denote $\frac{{\partial^2{l(\bm{\theta})}}}{\partial{\bm{\theta}}\partial{\bm{\theta}^{\mathrm{T}}}}=H(X_t,\bm{\theta})=(h_{ij}(X_t,\bm{\theta}))$ and $\mathrm{E}\frac{{\partial^2{l(\bm{\theta})}}}{\partial{\bm{\theta}}\partial{\bm{\theta}^{\mathrm{T}}}}=H(\bm{\theta})=(h_{ij}(\bm{\theta}))$. We only need to prove $h_{ij}(X_t,\bm{\theta})$ converges to a finite and non-stochastic function $h_{ij}(\bm{\theta})=\mathrm{E}(h_{ij}(X_t,\bm{\theta}))$. For $\mathrm{E}(h_{ij}(\bm{\theta})-\mathrm{E}(h_{ij}(X_t,\bm{\theta})))=0$, i.e.,
$$\sup_{\bm{\theta}\in \Theta}\left\|\frac{1}{n}\sum_{t=1}^n\dfrac{{\partial^2{l_t(\bm{\theta}^0)}}}{\partial{\bm{\theta}}\partial{\bm{\theta}^{\mathrm{T}}}}
-\mathrm{E}\dfrac{{\partial^2{l(\bm{\theta}^0)}}}{\partial{\bm{\theta}}\partial{\bm{\theta}^{\mathrm{T}}}}\right\|=o_p(1).$$
where $o_p(1)$ denote a random sequence converging to 0 in probability.\\
$\textbf{Step 4:}$ We check that $\bm{J}(\bm{\theta_0})=\mathrm{E}\left[\frac{\partial^2 l_t(\bm{\theta})}{\partial \bm{\theta}{\partial \bm{\theta}^{\mathrm{T}}}}\right]_{\bm{\theta_0}}$ is nonsingular. From the transition probability (\ref{transprob}), after some algebra, we have $\mathrm{E}\left(\frac{\partial^2}{\partial \bm{\theta}_i^2}\log P(x_{t-1},x_{t})\right)<\infty,~i=1,2,3$ and $\frac{\partial^2}{\partial \phi_1\partial \phi_2}\log P(x_{t-1},x_{t})=\frac{\partial^2}{\partial \phi_2\partial \phi_1} \log P(x_{t-1},x_{t})=0$, so we just need to prove that the following condition is true,
$$\mathrm{E}\left(\frac{\partial^2}{\partial \lambda\partial\phi_i}\log P(x_{t-1},x_{t})\right)=\mathrm{E}\left(\frac{\partial^2}{\partial \phi_i\partial \lambda}\log P(x_{t-1},x_{t})\right)< \infty,~i=1,2.$$
For convenience, we denote
\begin{align*}
p_1(x_{t-1},x_{t},\phi_{1},\lambda)
&=\sum_{m=0}^{\min(x_{t-1},x_t)}
\left(\begin{array}{cc}
i\\
m
\end{array}\right)
e^{-\lambda}\frac{\lambda^{j-m}}{(j-m)!}
\phi_1^m(1-\phi_1)^{i-m}\\
&=\sum_{m=0}^{\min(x_{t-1},x_t)}\rho_1(m,x_{t-1},x_{t},\phi_{1},\lambda)
\end{align*}
\begin{align*}
p_2(x_{t-1},x_{t},\phi_{2},\lambda)&=\sum_{m=0}^{x_{t}}
\frac{\Gamma(i+m)}{\Gamma(i)\Gamma(m+1)}\frac{\phi_2^m}{(1+\phi_2)^{i+m}}\frac{\lambda^{j-m}}{(1+\lambda)^{j-m+1}}\\
&=\sum_{m=0}^{x_{t}}\rho_2(m,x_{t-1},x_{t},\phi_{2},\lambda),
\end{align*}
we conclude that,
\begin{align*}
\frac{\partial \rho_1(m,x_{t-1},x_{t},\phi_{1},\lambda)}{\partial \phi_1}&=(\frac{m}{\phi_1}-\frac{x_{t-1}-m}{1-\phi_1}) \rho_1(m,x_{t-1},x_{t},\phi_{1},\lambda),\\
\frac{\partial \rho_2(m,x_{t-1},x_{t},\phi_{2},\lambda)}{\partial \phi_2}&=(\frac{m}{\phi_2}+\frac{x_{t-1}+m}{1+\phi_2}) \rho_2(m,x_{t-1},x_{t},\phi_{2},\lambda),\\
\frac{\partial \rho_1(m,x_{t-1},x_{t},\phi_{1},\lambda)}{\partial \lambda}&=(\frac{x_t-m}{\lambda}-1) \rho_1(m,x_{t-1},x_{t},\phi_{1},\lambda),\\
\frac{\partial \rho_2(m,x_{t-1},x_{t},\phi_{2},\lambda)}{\partial \lambda}&=(\frac{x_t-m}{\lambda}-\frac{x_t-m+1}{1+\lambda}) \rho_2(m,x_{t-1},x_{t},\phi_{1},\lambda).
\end{align*}
For $\frac{\partial^2}{\partial\phi_1\partial \lambda}\log P(x_{t-1},x_{t})$,
\begin{align}
&\frac{\partial^2}{\partial\phi_1\partial \lambda}\log P(x_{t-1},x_{t})=\frac{\partial^2}{\partial \lambda\partial\phi_1}\log P(x_{t-1},x_{t})=-\frac{1}{(P(x_{t-1},x_{t}))^2}\label{log_lphi}\\
&\times\left[\left(\sum_{m=0}^{x_{t}}(\frac{x_t-m}{\lambda}-1)\rho_1(m,x_{t-1},x_{t},\phi_{1},\lambda)\right) \left(\sum_{m=0}^{x_{t}}(\frac{m}{\phi_1}-\frac{x_{t-1}-m}{1-\phi_1})\rho_1(m,x_{t-1},x_{t},\phi_{1},\lambda)\right)\right]I_{1,t}^R\nonumber\\
&+\frac{1}{P(x_{t-1},x_{t})}\left(\sum_{m=0}^{x_{t}}(\frac{x_t-m}{\lambda}-1) (\frac{m}{\phi_1}-\frac{x_{t-1}-m}{1-\phi_1})\rho_1(m,x_{t-1},x_{t},\phi_{1},\lambda)\right)I_{1,t}^R.\nonumber
\end{align}
Note that
\begin{align*}
-\frac{x_{t-1}}{1-\phi_1}&\leq\frac{m}{\phi_1}-\frac{x_{t-1}-m}{1-\phi_1})\leq\frac{x_{t-1}}{\phi_1},\\
-1&\leq\frac{x_t-m}{\lambda}-1 <\frac{x_t}{\lambda}.
\end{align*}
Since $\phi_1\in(0,1)$ and $\lambda \in(0,\infty)$, it is easy to get
$$\mathrm{E}\left(\frac{\partial^2}{\partial \lambda\partial\phi_1}\log P(x_{t-1},x_{t})\right)=\mathrm{E}\left(\frac{\partial^2}{\partial \phi_1\partial \lambda}\log P(x_{t-1},x_{t})\right)< C_1\cdot \mathrm{E}X_t^2<\infty,(by~Proposition~\ref{momentexist}).$$
for some suitable constant $C_1$. By the same arguments as above, it follows that
$$\mathrm{E}\left(\frac{\partial^2}{\partial \lambda\partial\phi_2}\log P(x_{t-1},x_{t})\right)=\mathrm{E}\left(\frac{\partial^2}{\partial \phi_2\partial \lambda}\log P(x_{t-1},x_{t})\right)< C_2\cdot \mathrm{E}X_t^2<\infty,(by~Proposition~\ref{momentexist}).$$
for some suitable constant $C_2$.\\
From the above discussions, we can get
\begin{align*}
\sqrt{n}(\hat{\bm{\theta}}_{CML}-\bm{\theta_0})\overset{d}{\longrightarrow}N(\bm{0},\bm{J}^{-1}(\bm{\theta_0})\bm{I}(\bm{\theta_0})\bm{J}^{-1}(\bm{\theta_0})),
\end{align*}
where $\bm{I}(\bm{\theta_0})=\mathrm{E}\left[\frac{\partial l_t(\bm{\theta})}{\partial \bm{\theta}}\frac{\partial l_t(\bm{\theta})}{\partial \bm{\theta}^{\mathrm{T}}}\right]_{\bm{\theta_0}}$, $\bm{J}(\bm{\theta_0})=\mathrm{E}\left[\frac{\partial^2 l_t(\bm{\theta})}{\partial \bm{\theta}{\partial \bm{\theta}^{\mathrm{T}}}}\right]_{\bm{\theta_0}}$.
The proof of Theorem \ref{cmlzt} is completed.$\square$

{\bf \hspace{-1.6em}D-Ness algorithm}
We outline the precise procedures for applying the D-Ness algorithm to the BiNB-MTTINAR(1) model since we want to compare our suggested methods to the Ness algorithm. For a given $\lambda$, using standard least squares techniques, the sum of squared errors function is
\begin{equation*}
S_n(r,\lambda)=\sum_{t=1}^n\left(X_t-\sum_{k=1}^2\frac{\sum_{t=1}^nX_tX_{t-1}I_{k,t}(r)-\lambda\sum_{t=1}^nX_{t-1}I_{k,t}(r)}{\sum_{t=1}^nX_{t-1}^2 I_{k,t}(r)}\cdot X_{t-1} I_{k,t}(r)-\lambda\right)^{2},
\end{equation*}
where $I_{1,t}(r)=I\{{X_{t-1} \leq r}\},~I_{2,t}(r)=I\{{X_{t-1}>r}\}$ are both functions with respect to $r$.
Let
\begin{equation*}
J_n(r,\lambda)=\sum_{t=1}^n\left(X_t-\frac{\sum_{t=1}^nX_tX_{t-1}-\lambda\sum_{t=1}^nX_{t-1}}{\sum_{t=1}^nX_{t-1}^2}\cdot X_{t-1}-\lambda\right)^{2}- S_n(r,\lambda),
\end{equation*}
for a fixed $\lambda$, our aim is to estimate $r_{\lambda}$ by maximizing $J_n(r,\lambda)$, i.e.,
\begin{equation}\label{maxr}
\hat{r}_{\lambda}=\arg \max_{r \in [\underline{r},\overline{r}]}J_n(r,\lambda).
\end{equation}
Let $\underline{\lambda}$ and $\overline{\lambda}$ be some known lower and upper bounds of $\lambda$. We come to search for the final $\hat{r}_{CLS}^{(1)}$ by the following steps:
\begin{enumerate}[\bf \text{Step} 1.]
  \setlength{\itemsep}{1pt}
  \setlength{\parskip}{0pt}
  \setlength{\parsep}{0pt}
  \item Choose some appropriate positive integer $L$, let $\lambda(j)=\underline{\lambda}+\frac{j(\overline{\lambda}-\underline{\lambda})}{L}$,  $j=0,1,\cdots,L$.
  \item For each $j \in \{0,1,\cdots,L\}$, calculate $\hat{r}_{\lambda(j)}$ by (\ref{maxr}).
  \item The final $\hat{r}_{CLS}^{(1)}$ is calculated by $
\hat{r}_{CLS}^{(1)}=\arg \max_{0 \leq j \leq L}J_n(\hat{r}_{\lambda(j)},\lambda(j)).\nonumber
$ \hspace{1.36in}$\square$
\end{enumerate}

{\bf \hspace{-1.6em}Proof of Theorems \ref{consist-var}}
From he first partial derivative equal to 0, after some calculations, we obtain $\hat{\bm{\vartheta}}$ closed-form expressions as follows
\begin{align*}
\hat{\sigma}_{1}^2&=\frac{\sum_{t=1}^{n}I_{1,t}^R\sum_{t=1}^n I_{1,t}^RV_t X_{t-1}-\sum_{t=1}^{n}I_{1,t}^RX_{t-1}\sum_{t=1}^{n}I_{1,t}^RV_t}{\sum_{t=1}^{n}I_{1,t}^R\sum_{t=1}^{n}I_{1,t}^RX_{t-1}^2  - (\sum_{t=1}^{n}I_{1,t}^RX_{t-1})^2},\\
\hat{\sigma}_{2}^2&=\frac{\sum_{t=1}^{n}I_{2,t}^R\sum_{t=1}^{n}I_{2,t}^RV_t X_{t-1}-\sum_{t=1}^{n}I_{2,t}^RX_{t-1}\sum_{t=1}^{n}I_{2,t}^RV_t}{\sum_{t=1}^{n}I_{2,t}^R\sum_{t=1}^{n}I_{2,t}^RX_{t-1}^2- (\sum_{t=1}^{n}I_{2,t}^RX_{t-1})^2},\\
\hat{b}_{1}&=\frac{\sum_{t=1}^{n}I_{1,t}^RX_{t-1}^2\sum_{t=1}^{n}I_{1,t}^RV_t-\sum_{t=1}^{n}I_{1,t}^RX_{t-1}\sum_{t=1}^n I_{1,t}^RV_t X_{t-1}}{\sum_{t=1}^{n}I_{1,t}^R\sum_{t=1}^{n}I_{1,t}^RX_{t-1}^2  - (\sum_{t=1}^{n}I_{1,t}^RX_{t-1})^2},\\
\hat{b}_{2}&=\frac{\sum_{t=1}^{n}I_{2,t}^RX_{t-1}^2\sum_{t=1}^{n}I_{2,t}^RV_t-\sum_{t=1}^{n}I_{2,t}^RX_{t-1}\sum_{t=1}^{n}I_{2,t}^RV_t X_{t-1}}{\sum_{t=1}^{n}I_{2,t}^R\sum_{t=1}^{n}I_{2,t}^RX_{t-1}^2- (\sum_{t=1}^{n}I_{2,t}^RX_{t-1})^2}.
\end{align*}
From the properties of Bernoulli, Geometric, Poisson random variables, there is
\begin{align*}
&{\rm E}(B_i-{\rm E}(B_i))^4=(1-\phi_1)^3\phi_1+3(1-\phi_1)^2\phi_1^2<\infty,\\
&{\rm E}(W_i-{\rm E}(W_i))^4=\phi_2+10\phi_2^2+18\phi_2^3+8\phi_2^4<\infty,\\
&{\rm E}(Z_{1,t}-{\rm E}(Z_{1,t}))^4=\lambda+3\lambda^2<\infty,\\
&{\rm E}(Z_{2,t}-{\rm E}(Z_{2,t}))^4=\lambda+10\lambda^2+18\lambda^3+8\lambda^4<\infty.
\end{align*}
It follows from Theorems 3.3 in \cite{Win1991} that $\hat{\bm{\vartheta}}$ is a consistent and asymptotically normal estimator of $\bm{\vartheta}$. The proof of Theorem \ref{consist-var} is completed. \hspace{2.36in}$\square$



\end{document}